\documentclass[
 aps,
 11pt,
 final,
 notitlepage,
 oneside,
 onecolumn
 nobibnotes,
 nofootinbib,
 superscriptaddress,
 noshowpacs,
 centertags]
{revtex4}

\def\saoname{Special Astrophysical Observatory,  Russian Academy of Sciences,
              Nizhnii Arkhyz, 369167 Russia}

\def\squareforqed{\hbox{\rlap{$\sqcap$}$\sqcup$}}

\def\sq{\ifmmode\squareforqed\else{\unskip\nobreak\hfil
\penalty50\hskip1em\null\nobreak\hfil\squareforqed
\parfillskip=0pt\finalhyphendemerits=0\endgraf}\fi}

\def\degr{\hbox{$^\circ$}}

\def\arcmin{\hbox{$^\prime$}}

\def\arcsec{\hbox{$^{\prime\prime}$}}

\def\utw{\smash{\rlap{\lower5pt\hbox{$\sim$}}}}

\def\udtw{\smash{\rlap{\lower6pt\hbox{$\approx$}}}}

\def\fm{\hbox{$\,.\!\!^{\rm m}$}}

\def\fdg{\hbox{$\,.\!\!^\circ$}}

\def\farcm{\hbox{$\,.\mkern-4mu^\prime$}}

\def\farcs{\hbox{$\,.\!\!^{\prime\prime}$}}

\def\diameter{{\ifmmode\mathchoice
{\ooalign{\hfil\hbox{$\displaystyle/$}\hfil\crcr
{\hbox{$\displaystyle\mathchar"20D$}}}}
{\ooalign{\hfil\hbox{$\textstyle/$}\hfil\crcr
{\hbox{$\textstyle\mathchar"20D$}}}}
{\ooalign{\hfil\hbox{$\scriptstyle/$}\hfil\crcr
{\hbox{$\scriptstyle\mathchar"20D$}}}}
{\ooalign{\hfil\hbox{$\scriptscriptstyle/$}\hfil\crcr
{\hbox{$\scriptscriptstyle\mathchar"20D$}}}}
\else{\ooalign{\hfil/\hfil\crcr\mathhexbox20D}}%
\fi}}

\newcommand{\ab}{Astrophysical Bulletin }
\newcommand{\aap}{Astron. and Astrophys. }
\newcommand{\aaps}{Astron. and Astrophys. Suppl. }
\newcommand{\aj}{Astron.~J. }
\newcommand{\apjs}{Astrophys.~J. Suppl. }
\newcommand{\mnras}{Monthly Notices Royal Astron. Soc. }
\newcommand{\alet}{Astronomy Letters }
\begin{document}
\selectlanguage{english}

\title{The Excess Density of Field Galaxies near $\boldsymbol{z\sim 0.56}$ around 
the Gamma-Ray Burst GRB\,021004 Position}

\author{\firstname{I.~V.}~\surname{Sokolov}}
\email{sok334455@mail.ru}
    \affiliation{Institute of Astronomy, Russian Academy of Sciences, Moscow, 119017 Russia}

\author{\firstname{A.~J.}~\surname{Castro-Tirado}}
    \affiliation{Stellar Physics Department, Institute for Astrophysics of Andalucia (IAA-CSIC), Granada, 18008 Spain}

\author{\firstname{O.~P.}~\surname{Zhelenkova}}
    \affiliation{\saoname}
    \affiliation{ITMO University, St.~Petersburg, 197101 Russia}
\author{\firstname{I.~A.}~\surname{Solovyev}}
    \affiliation{Astronomical Department, St.~Petersburg State University, St.~Petersburg, 199034 Russia}

\author{\firstname{O.~V.}~\surname{Verkhodanov}}
    \affiliation{\saoname}

\author{\firstname{V.~V.}~\surname{Sokolov}}
        \affiliation{\saoname}

\begin{abstract}
We test for reliability any signatures of field galaxies clustering in
the GRB\,021004 line of sight.
The first signature is the GRB\,021004 field photometric redshifts 
distribution based on the 6-m telescope of the Special Astrophysical
Observatory of the Russian Academy of Sciences observations
with a peak near $z\sim0.56$ estimated from multicolor 
photometry in the GRB direction. The second signature is the 
Mg\,II$\lambda\lambda2796,2803$\r{A}\r{A} absorption doublet at $z\approx0.56$ 
in VLT/UVES spectra obtained for the GRB\,021004 afterglow. The third
signature is the galaxy clustering in a larger
(of about $3\degr\times3\degr$)
area around GRB\,021004 with an effective peak near $z\sim0.56$ for
both the spectral and photometric redshifts from a few
catalogs of clusters based on the Sloan Digital Sky Survey (SDSS)
and Baryon Oscillation Spectroscopic Survey (BOSS) as a part of
SDSS-III. From catalog data the size of the whole inhomogeneity in
distribution of the galaxy clusters with the peak near $z\approx0.56$ 
is also estimated as about $6\degr$--$8\degr$ or
140--190~Mpc. A possibility of inhomogeneity (a galaxy cluster)
near the GRB\,021004 direction can be also confirmed by an
inhomogeneity in the cosmic microwave background related with the
Sunyaev--Zeldovich effect.
\end{abstract}

\maketitle

\section{introduction}
The main motivation in
conducting this work is a number of additional new studies on the
Great Walls in increasing order of redshift  $z$:
\begin{list}{}{
\setlength\leftmargin{5mm} \setlength\topsep{1mm}
\setlength\parsep{-0.5mm} \setlength\itemsep{2mm} }
 \item[$\bullet$] Paper~\cite{1:Sokolov_en} deals with 
the Sloan Great Wall superclusters in
the redshift range $0.04<z<0.12$. The authors explore how
unusual are the Shapley Supercluster (with $z=0.046$) and the
Sloan Great Wall, considering the Sloan Great Wall as a complex of
superclusters with collapsing cores~\cite{2:Sokolov_en}. In
these papers the distribution of galaxy groups in the Sloan Great
Wall superclusters in the sky plane
is studied in detail.%
 \item[$\bullet$] The BOSS Great Wall superclusters in the
redshift range $0.43<z<0.71$ were studied in
paper~\cite{3:Sokolov_en}, it deals with discovery of a massive
supercluster system at $z\sim0.47$. The BOSS Great Wall
consists of two walls with diameters about 186 and  173\,Mpc,
and two other major superclusters with diameters of about  64
and 91\,Mpc. This system consists of 830 galaxies with 
the mean $z=0.47$ and with the total mass 
approximately~\mbox{$2\times10^{17} M_\odot$}.  The authors emphasize 
that the morphology of the superclusters in the BOSS Great Wall system 
is similar to the morphology of the superclusters in the Sloan Great Wall.
 \item[$\bullet$] Paper~\cite{4:Sokolov_en} concerns some new 
Large Quasar Group (Huge-LQG) at $z\sim1.3$. It was the study 
of sky distribution of 73 quasars ($z=1.27$) of the new Huge-LQG, 
together with that of 34~quasars of the Clowes-Cappusano Large Quasar Group
(CCLQG) 
with \mbox{$z=1.28$}~\cite{37:Sokolov_en}. 
``Huge-LQG'' is the area of about \mbox {$29\fdg5\times24\fdg0$}. 
The members of each LQG are connected at the linkage scale of 100~Mpc.
 \item[$\bullet$] Here it should be said also about a giant ring-like structure
at $0.78<z<0.86$ displayed by GRBs themselves, which is discussed
in paper~\cite{6:Sokolov_en}. This appears to indicate the
presence of some real density enhancement at about 2800~Mpc.  The
possible structure in the GRB sky distribution is also discussed
in~\cite{41:Sokolov_en}. This huge GRB structure lies ten times
farther away than the Sloan Great Wall, at a distance of
approximately ten billion light years. The size of the structure
defined by these GRBs is about 2000--3000\,Mpc, or six times
larger than the size of the Sloan Great Wall.%
 \item[$\bullet$] Paper~\cite{7:Sokolov_en} deals with statistics 
of the Planck CMB signal in direction of GRBs from the BATSE 
and {\it BeppoSAX} catalogs or on a possible inhomogeneity (non-Gaussianity) 
in the GRB sky distribution. So, although the apparent distribution of GRBs in
the sky was proven to be isotropic~\cite{8:Sokolov_en} some
studies have pointed out to a possible correlation with a
nonuniformity of the cosmic microwave background
(CMB)~\cite{7:Sokolov_en} and clustering of GRBs at intermediate
redshifts has been also reported~\cite{6:Sokolov_en}. Besides
this, a possible clustering of neutrino signals in relation with
GRBs has been also discussed~\cite{9:Sokolov_en}.
\end{list}

The Mg\,II $\lambda\lambda 2796,2798$\r{A}\r{A} absorbtion doublet in the
joint redshift range of the above-mentioned Huge-LQG  (Large
Quasar Group at  $z \sim 1.3$) has been found in spectra of
background quasars (with \mbox{$z \gtrsim 1.4$}),  using the
Mg\,II absorber catalogue of Raghunathan et al. of
2016~\cite{11:Sokolov_en}. Another work~\cite{12:Sokolov_en}
presents statistics from a survey for intervening Mg~II absorption
towards 100 quasars with emission (proper) redshifts between
\mbox{$z=3.55$} and $z=7.08$. 
So, the quasar spectroscopy relies on using some
strong detecting galaxy clusters.
So, GRB afterglow spectroscopy also can reveal
intervening systems along the line of sight. The clustering of
galaxies and the clustering around GRB sight-lines are detected
and studied in the same manner employed with quasar spectroscopy
in many papers already~(see~\cite{10:Sokolov_en} and references
therein)  because GRBs originate at cosmological distances with
energy releases of $10^{51}$--$10^{53}$~erg at a range of
redshifts between about 0.01 and  more than  9.2.  And
long-duration GRBs are beacons of starforming galaxies (see
references in~\cite{42:Sokolov_en}) up to very high redshifts.

Galaxies that give rise to the absorption line systems in GRBs
afterglow spectra have been directly imaged and investigated. 
So, it is possible to try to find the overdensity excess of GRB field
galaxies around GRB positions by different methods: spectroscopic,
photometric ones (using deep images and multiband photometry) plus
the search for a correlation with Cosmic Microwave Background (CMB).
The data from new redshift catalogs of galaxy clusters can be used
also to estimate size of these galaxy superclusters in regions near
the GRB positions. In particular, in this paper we test for
reliability any signatures of field galaxies clustering in
GRB\,021004 line of sight and in the larger areas around the GRB.

\section{ON THE OBSERVATIONAL DATA AND METHODOLOGY OF THEIR INTERPRETATION}
\label{sec:on_the_observational_data_and_methodology:Sokolov_en}
Here the 6-m telescope of the Special Astrophysical Observatory (BTA)
observational data are the base of our investigations. The deep fields
reported in Table~\ref{Tab:fields:Sokolov_en} were studied with BTA (before
2002) under the program of GRBs monitoring, starting from the very
first optical identifications of their GRB afterglows. Hereby, all
these BTA observations were carried out in the course of the search
and study of GRB host galaxies and comparison of their properties
with those of all observable galaxies in GRB deep fields.
Papers~\cite{13:Sokolov_en,14:Sokolov_en} describe the methodology of processing and
interpretation of these observations, which can be used now in new
tasks on the study of distribution of galaxy clusters in GRB
directions~\cite{10:Sokolov_en,18:Sokolov_en,19:Sokolov_en,20:Sokolov_en}.
\begin{table}
    \setcaptionmargin{0mm} \onelinecaptionstrue \captionstyle{normal}
    \caption{Deep fields around gamma-ray bursts observed with BTA in 1987--2002}
    \label{Tab:fields:Sokolov_en}
    \medskip
    \begin{tabular}{c|c|c}
        \hline
        GRB & Bands & Exposure time, s \\
        \hline
        970508 & $BVRI$ & $ 600 \times 7, 500 \times 4, 600 \times 5, 400 \times 5 $ \\
        971214 & $VR$   & $ 600 \times 1, 600 \times 1 $ \\
        980613 & $BVRI$ & $ 700 \times 1, 600 \times 1, 600 \times 3 $ \\
        980703 & $BVRI$ & $ 480 \times 1, 320 \times 1, 300 \times 1, 360 \times 1 $ \\
        990123 & $BVRI$ & $ 600 \times 1, 600 \times 1, 600 \times 1, 600 \times 1 $ \\
        991208 & $BVRI$ & $ 300 \times 6, 300 \times 5, 180 \times 7, 180 \times 2 $ \\
        000926 & $BVRI$ & $ 500 \times 5, 300 \times 5, 180 \times 25, 120 \times 15 $ \\
        021004 & $BVRI$ & $ 600 \times 6, 450 \times 13, 180 \times 15, 120 \times 14 $ \\
        \hline
    \end{tabular}
\end{table}
\begin{figure*}
 \onelinecaptionstrue \captionstyle{normal} \setcaptionmargin{5mm}
 \includegraphics[width=0.95\columnwidth]{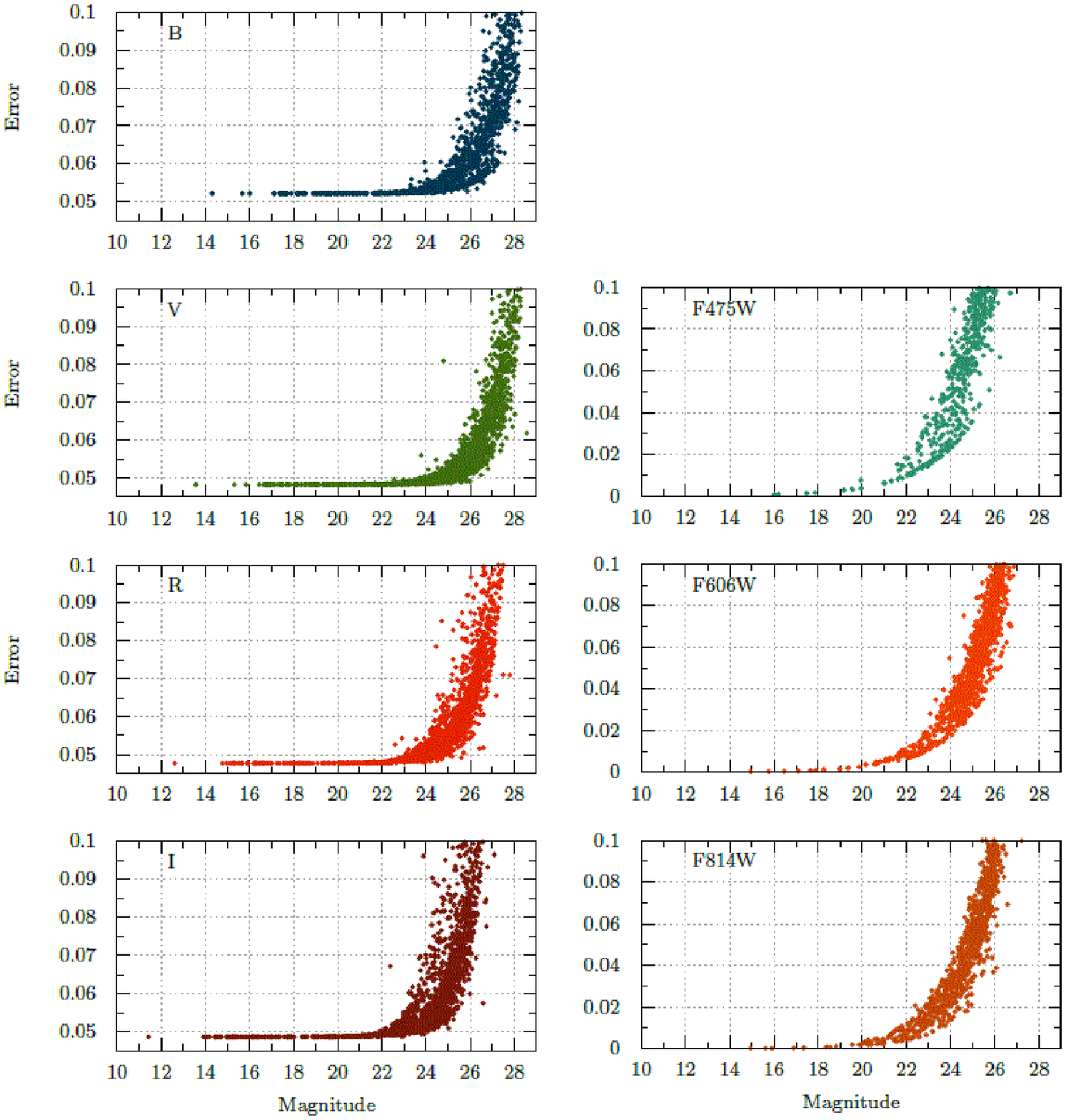}
 \caption{The magnitude-error diagram for objects extracted from 
the GRB\,021004 field. Left column graphs represent data of BTA $BVRI$ bands. 
Right column graphs represent data for the same field acquired with 
HST~ACS camera in $F475W, F606W$ and $F814W$ optical bands 
(Hubble Legacy Archive, {\tt http://hla.stsci.edu/hlaview.html}).}
  \label{mag-err:Sokolov_en}
\end{figure*}
In particular, to study the GRB\,021004 field we used the methodology
already developed for the $BVRI$ deep field of GRB\,000926 (see
Table~\ref{Tab:fields:Sokolov_en}), which was investigated with 
BTA and the results were published~\cite{13:Sokolov_en}.
This work presents the observations
of a $3\farcm6\times3\arcmin$ field centered on the
GRB\,000926 host galaxy position at coordinates 
\mbox{RA(J2000)$=17^{\rm h}04^{\rm m} 11^{\rm s}$},
\mbox{Dec(J2000)$=+51\degr4\arcmin9\farcs8$}.
Hereafter notations RAJ and DecJ are used.
The observations were carried out on the BTA using the SCORPIO
instrument~\cite{44:Sokolov_en}.
\begin{figure}
    \onelinecaptionstrue \captionstyle{normal} \setcaptionmargin{5mm}
    \includegraphics[width=1\columnwidth]{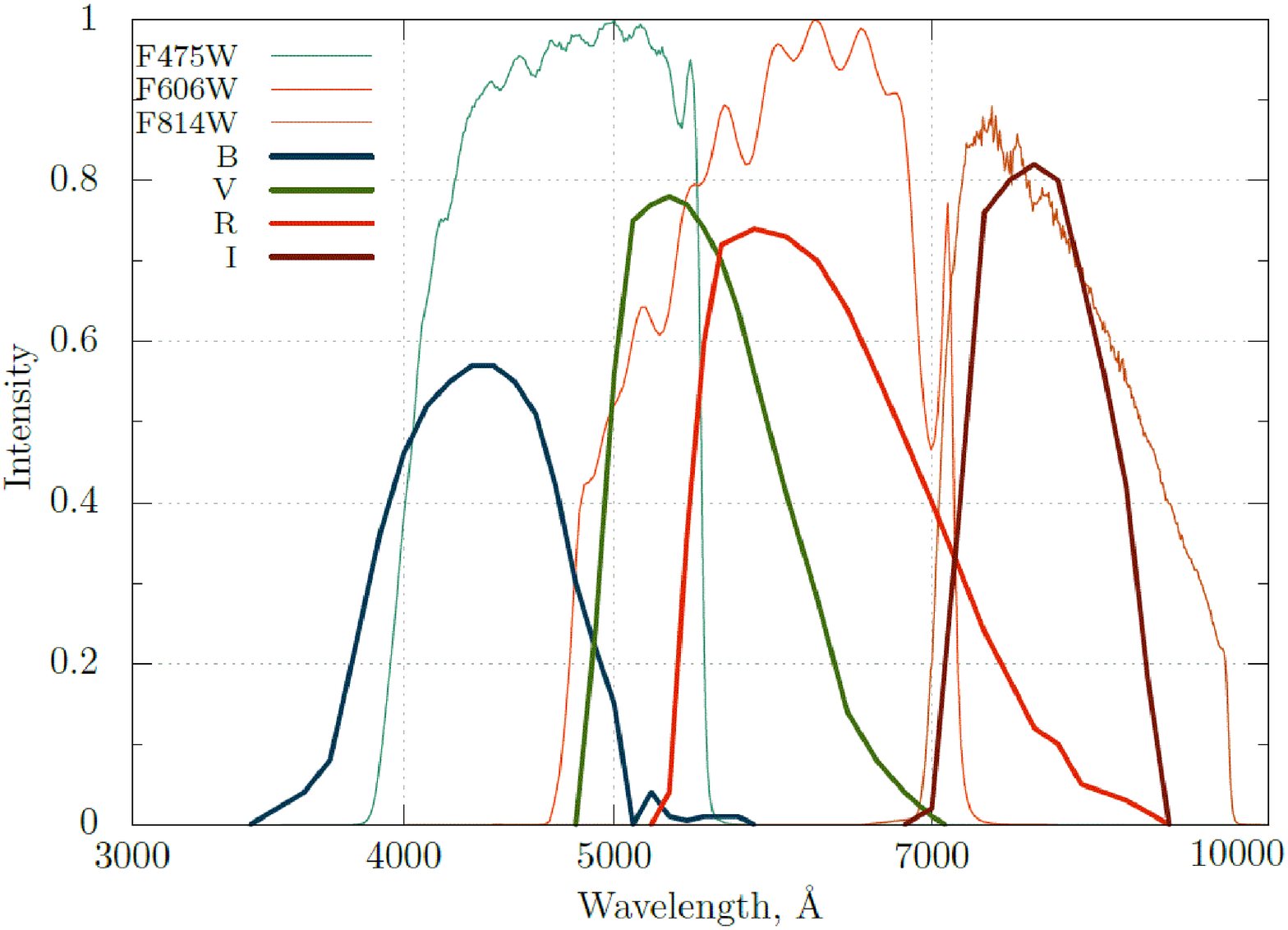}
    \caption{Filter transmission curves of BTA $BVRI$ and HST~ACS 
             ($F475W$, $F606$, $F814W$) optical bands.}
    \label{bands:Sokolov_en}
\end{figure}
The catalog of galaxies detected in this GRB\,000926 field
includes 264 objects, for which the signal-to-noise ratio is larger
than 5 in each photometric band.  The following limiting magnitudes
were derived: $26\fm6$~($B$), $25\fm0$~($V$), $25\fm8$~($R_c$), and
$24\fm5$~($I_c$). The differential galaxy counts are in good
agreement with previously published CCD observations of the deep
fields. The photometric redshifts for all of the cataloged objects
and the corresponding color variations of the galaxies as a function
of the redshift were derived. As a result, for luminous spiral
galaxies with $M_{B}<18$, no evidence for any noticeable evolution of
their linear sizes to $z\sim1$ was found.

For studying other deep fields listed in Table~\ref{Tab:fields:Sokolov_en} 
these data were supplemented with other data obtained with other 
instruments for GRB\,021004, GRB\,970508, and others, including
investigation of the surroundings of the radio source
RC~J0311+0507~\cite{43:Sokolov_en}.

And as was mentioned above , the GRB\,021004 field was studied as a
part of GRB afterglow observations
program~\cite{13:Sokolov_en,14:Sokolov_en}. Exposure times for
this field were of about one hour in each of the $BVRI$ optical
bands (see Table~\ref{Tab:fields:Sokolov_en}). We used the
SExtractor software package~\cite{15:Sokolov_en} to extract
objects from the stacked $BVRI$ image. The catalog of galaxies,
extracted from this field of size $4\arcmin \times 4\arcmin$
includes 935 objects for which magnitude errors are not larger
than $0\fm1$ (see Fig.~\ref{mag-err:Sokolov_en}, left). So, the
following limiting magnitudes were achieved for these BTA
observations of the GRB\,021004 field: $26\fm9$~($B$),
$27\fm2$~($V$), $26\fm0$~($R_c$) and $25\fm5$~($I$).

\begin{figure}
\onelinecaptionstrue \captionstyle{normal} \setcaptionmargin{5mm}
 \includegraphics[width=1\columnwidth]{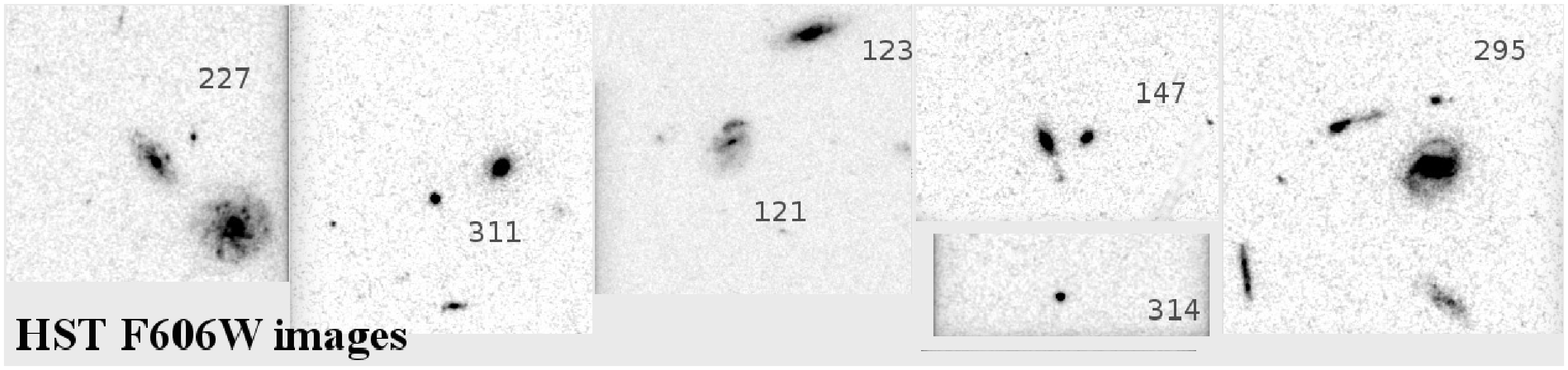}
    \caption{Images of seven galaxies extracted from HST~ACS $F606W$ 
    (HST Legacy Archive, {\tt{http://hla.stsci.edu/hlaview.html}})
    near the GRB\,021004 position within the field of 
    size $3\farcs5\times3\farcs5$. The object numbers from 
    the HST~ACS catalog are indicated. The table~\ref{tab:emb:Sokolov_en} 
    contains the estimates of photometric $z$, with the probability of 
    the photo-$z$ measurements for these seven galaxies with known 
    morphology, see the text.}
    \label{HST-galaxies:Sokolov_en}
\end{figure}

One of the goals of this work is to point out the possibility of
using the 6-m telescope in these challenging and actual tasks to
supplement mentioned studies with other instruments' data. We also
made sure our results of deep photometry are consistent with the
corresponding HST~ACS data\footnote{Hubble Legacy Archive,
\url{http://hla.stsci.edu/hlaview.html}} for the GRB\,021004
field. Figure~\ref{mag-err:Sokolov_en} (right)  represents data
acquired with HSC ACS camera in $F475W$, $F606W$ and $F814W$
optical bands with the magnitude errors no larger than 0.1 for the
same extended objects, whose $BVRI$ photometry was made with BTA.
Figure~\ref{bands:Sokolov_en} shows the filter transmission
curves of BTA $BVRI$ optical bands and also HST~ACS optical bands
$F475W$, $F606W$, $F814W$ for comparison. Here we did not use
directly these HST~ACS data (Fig.~\ref{mag-err:Sokolov_en},
right) for the photometric redshift estimations, since the filter
transmission curves of BTA $BVRI$ and the HST~ACS optical bands
overlap. But the main thing is that the same HST~ACS data can be
used also for obtaining information on morphology of HST/BTA
objects (see Fig.~\ref{HST-galaxies:Sokolov_en}).

\begin{table}
        \setcaptionmargin{0mm} \onelinecaptionstrue \captionstyle{normal}
        \caption{The estimates of photometric $z$, with the probability of the photo-$z$
measurements for the seven galaxies with known morphology near the GRB 021004 
position (see Fig.~\ref{HST-galaxies:Sokolov_en} and the text) }
        \label{tab:emb:Sokolov_en}
        \medskip
        \begin{tabular}{c|c|c|c|c}
                \hline
                Num  & $R$, mag & $z$   & \% & Type  \\
                \hline
                227  & 20.60    & 0.435 & 82 & S0    \\
                311  & 21.82    & 0.40  & 94 & Burst \\
                121  & 21.16    & 0.42  & 84 & Sa    \\
                123  & 21.41    & 0.40  & 98 & E     \\
                147  & 21.61    & 0.44  & 86 & E     \\
                314  & 21.58    & 0.46  & 95 & Burst \\
                295  & 20.60    & 0.41  & 92 & S0    \\
                \hline
        \end{tabular}
\end{table}

The HST photometry of extended objects in the GRB\,021004 field is
adduced here for comparison with data of BTA $BVRI$ photometry of
the same objects. In particular, as is seen in
Fig.~\ref{mag-err:Sokolov_en}, the magnitude--error diagram for
objects extracted from the GRB\,021004 field in BTA ($R$, $I$)
(left) and HST~ACS ($F606W, F814W$) (right) optical bands agree
satisfactorily. This corresponds to the fact that the transmission
curves of the filters mentioned above  are the closest to each
other in central wavelengths, half-widths and maxima
(Fig.~\ref{bands:Sokolov_en}).

Besides studying host galaxies of GRBs, prompt spectroscopy of
their afterglows has revealed (in the optical) such intervening
absorption systems as found in the past with quasar spectroscopy.
One of such systems is Mg~II ($\lambda\lambda2796,2798$\r{A}\r{A} at rest
frame), which is strong and ease to detect in moderate $S/N$
spectra. Using a large sample of GRB afterglows, an overdensity
(the factor 2--4) of strong absorption line systems along the
lines of sight has been
found~\mbox{\cite{18:Sokolov_en,19:Sokolov_en,20:Sokolov_en}}.
We use these data in this work for direct confirmation of the
overdensity excess of field galaxies with the peak near $z\sim0.56$ 
around the GRB\,021004 position.

Along with photometric and spectral methods in studies of GRB fields, 
we also include photometry of clusters in new catalogs of galaxy clusters with
the purpose to confirm such a peak in $z$ and to estimate angular
size of the whole inhomogeneity in distribution of galaxy
clusters, i.e. superclusters at $z\sim0.56$.
\begin{table}
    \setcaptionmargin{0mm} \onelinecaptionstrue \captionstyle{normal}
    \caption{The absorption Mg\,II $\lambda\lambda2796,2803$\r{A}\r{A} doublet 
    identified in~\cite{19:Sokolov_en} in spectra of GRB\,021004 }
    \label{Tab:MgIIlines:Sokolov_en}
    \medskip
    \begin{tabular}{c|c|c|c}
        \hline
	GRB         & $z_{\rm GRB}$ & $z_{\rm abs}$ & $W_r(\lambda 2796$\r{A}), \r{A} \\
        \hline
                    &               & 0.5550        & $0.248 \pm 0.025$ \\
        021004      & 2.3295        & 1.3800        & $1.637 \pm 0.020$ \\
                    &               & 1.6026        & $1.407 \pm 0.024$ \\
        \hline
    \end{tabular}
 \end{table}

\section{THE PHOTOMETRIC REDSHIFT DISTRIBUTION OF GRB\,021004 FIELD
GALAXIES AND THE Mg~II $\lambda\lambda2796,2803$\r{A}\r{A} ABSORPTION DOUBLET
AT $z \approx 0.56$ IN THE GRB\,021004 AFTERGLOW SPECTRUM } 
\label{sec:redshift_distribution:Sokolov_en}
The main idea of the photometric
redshift estimations is as follows~\cite{21:Sokolov_en}: the object multicolor
photometry may be considered as some very low-resolution spectrum
(more specifically, as an energy distribution in several bands),
which is used to define redshifts for many (up to several hundreds)
objects at once.

In practice, we estimated the photometric redshifts for the extended
objects of our BTA sample using the Hyperz software package -- a
public photometric redshift code~\cite{22:Sokolov_en}. The method
is based on finding the best fit of template spectra of various
galaxy types, used in the photometric redshift calculation. The
template spectra used in our photometric redshift calculation were
also taken from~\cite{22:Sokolov_en}.
\begin{figure}
    \onelinecaptionstrue \captionstyle{normal} \setcaptionmargin{5mm}
    \includegraphics[width=1\columnwidth]{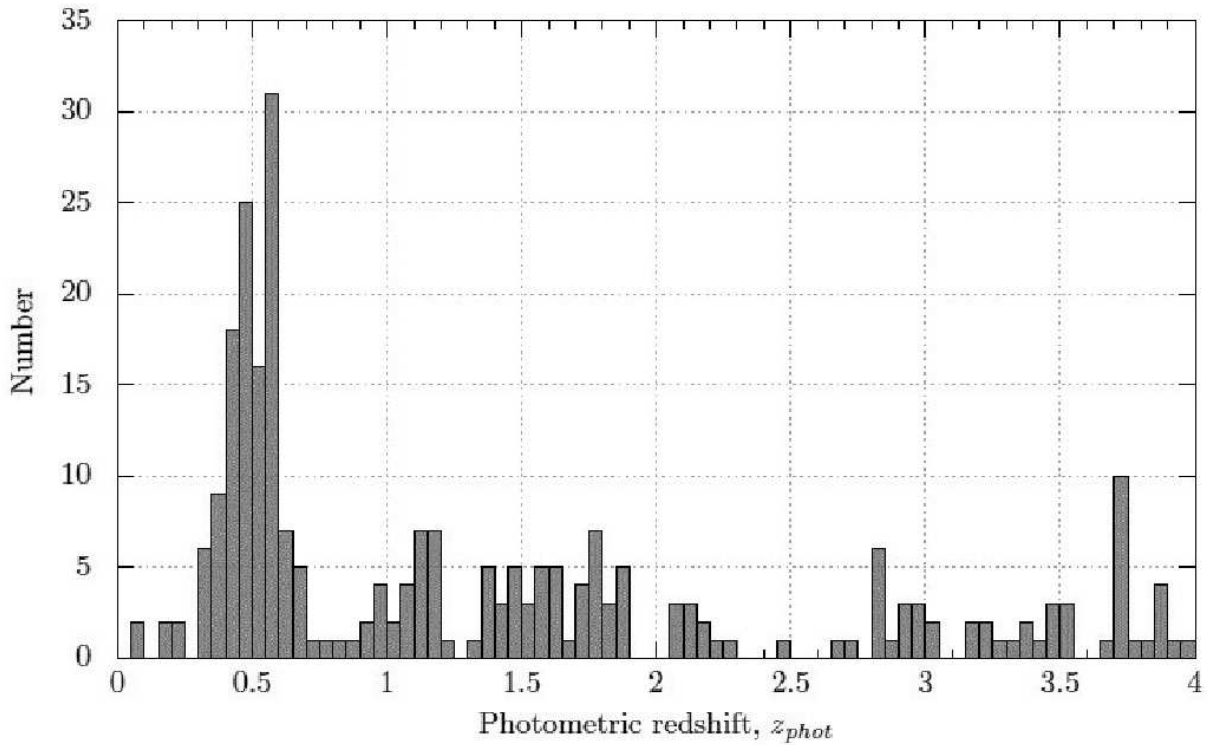}
    \caption{The photometric redshift distribution for 246 objects with 
    the peak at $z\approx0.56$ based on BTA $BVRI$ data.}
    \label{z-distribution:Sokolov_en}
\end{figure}
The input data for Hyperz were: the apparent magnitudes of the
objects in four $BVRI$ bands, the internal extinction law and the
redshift range, in which the solution is applied (we considered
$z$ from~0 to~4).  We used the extinction law provided by Calzetti
et al.~\cite{23:Sokolov_en} for starburst galaxies, which is
most commonly used for studies similar to our own as in the
above-mentioned work~\cite{16:Sokolov_en}, where the application
of the Hyperz method is described in more detail and which was
applied there for interpretation of observational BTA data.

For seven bright galaxies (with \mbox{$R=20\fm6$--$21\fm8$}) in the
BTA GRB\,021004 field we found that models using the spectra
from~\cite{22:Sokolov_en} and assigned to these galaxies are in
good agreement with HST~ACS data. The later are directly revealing the
morphology of these objects thanks to the better angular
resolution of the HST images (see Fig.~\ref{HST-galaxies:Sokolov_en}).
Table~\ref{tab:emb:Sokolov_en} contains the estimated probability
of the photometric redshift $z\approx0.4$ thus showing the
importance of the initial model spectra or template spectra (or
spectral energy distributions---SEDs) of various galaxy types, when
the morphology of galaxies from BTA observations can be determined
only by their spectral energy distributions for fainter objects, etc.
(The standard SEDs for the Hyperz fitting procedures can be taken
from \url{http://webast.ast.obs-mip.fr/hyperz}~\cite{22:Sokolov_en}).
\begin{figure}
    \onelinecaptionstrue \captionstyle{normal} \setcaptionmargin{5mm}
    \includegraphics[width=1\columnwidth]{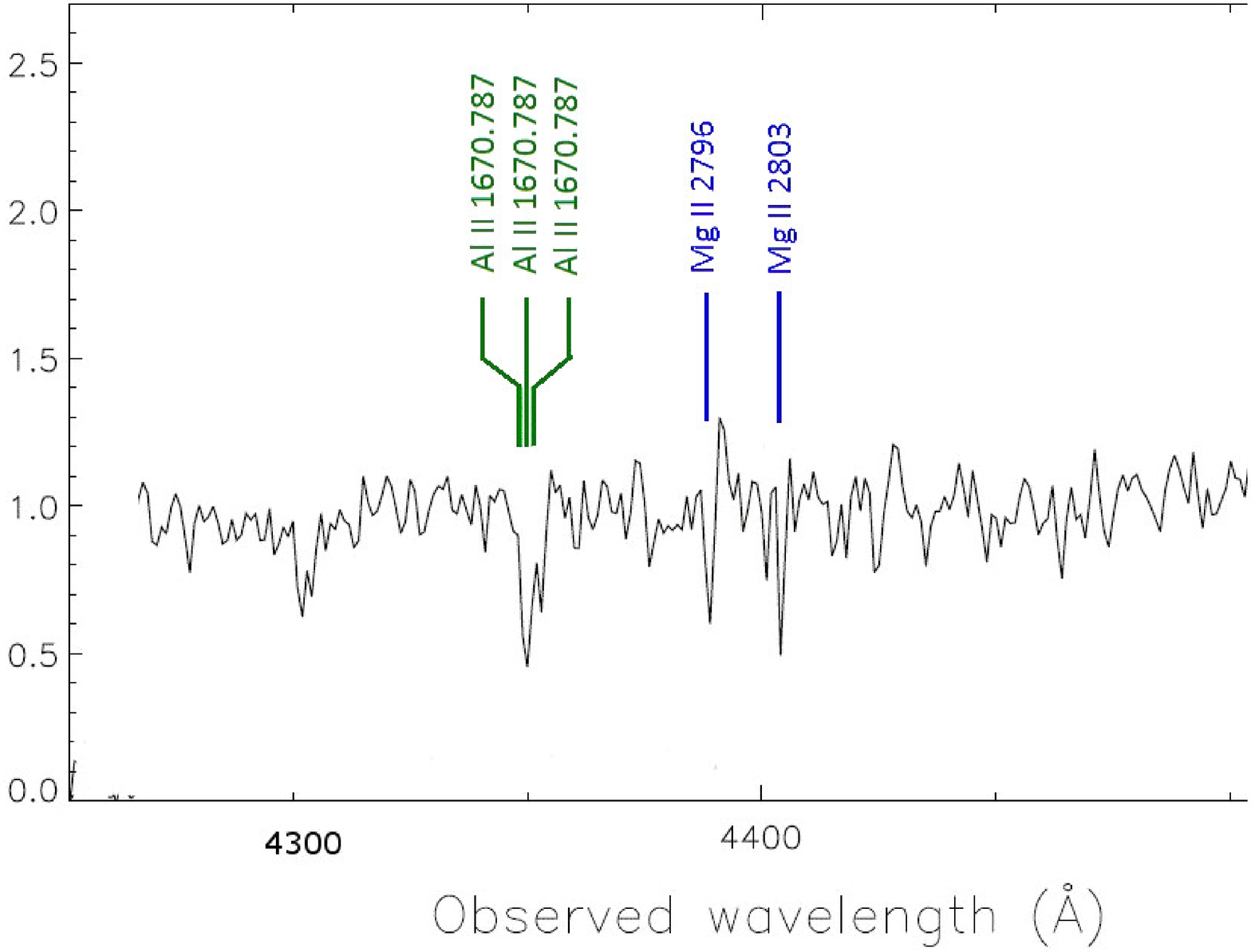}
    \caption{Features of VLT/UVES spectra near 4400\r{A} with absorption  
 could be interpreted as Mg\,II~doublet redshifted at $z\approx0.56$.}
    \label{spectra:Sokolov_en}
\end{figure}
So, using BTA observations of the GRB\,021004 field in $BVRI$ bands
for the spatial (redshift) distribution of field galaxies, we have found a
large inhomogeneity with a maximum in the photometric redshift
distribution (see Fig.~\ref{z-distribution:Sokolov_en}) at $z \approx 0.56$ for
246 objects (from 935~objects, Fig.~\ref{mag-err:Sokolov_en}) in the GRB field
with the probability of redshift measurements~\cite{16:Sokolov_en} about 90\%
and more (up to 99\%). 
The difference of these redshifts estimates,
for $z<1$ at least, from spectral \textit{z} is about 10\% (see the
next section), which is high enough for statistical study of objects
properties (more exactly, here we are interested in redshifts of
objects with \mbox{$0<z<0.7$}; see below).

Certainly, comparison of our data with those of HST
(Fig.~\ref{mag-err:Sokolov_en} and Fig.~\ref{HST-galaxies:Sokolov_en}) 
is not yet a significant test of redshift determination near 
the peak $z\approx0.56$ in Fig.~\ref{z-distribution:Sokolov_en}. 
That is why below we also use VLT/UVES spectral observations of this GRB
afterglow. The GRB\,021004 afterglow spectrum has been measured by
two teams~\cite{19:Sokolov_en,20:Sokolov_en}. The spectrum
from~\cite{20:Sokolov_en} shows evidence of Mg\,II$\lambda\lambda2796,2803$\r{A}\r{A} 
doublet redshifted at $z=1.3820$ and \mbox{$z=1.6020$} 
(see Fig.~\ref{mag-err:Sokolov_en} in~\cite{20:Sokolov_en}). 
This spectrum also includes two features that could be identified as 
Mg\,II doublet (in the GRB\,021004 line of sight) redshifted at 
\mbox{$z\approx0.5550$--$0.5570$}: see Fig.~\ref{spectra:Sokolov_en}
from~\cite{20:Sokolov_en} and results of identification 
on the earlier spectrum of this burst from~\cite{19:Sokolov_en} in
Table~\ref{Tab:MgIIlines:Sokolov_en}.

\section{ON THE GALAXY CLUSTERING WITH AN EFFECTIVE PEAK NEAR $z \sim 0.56$ 
FOR THE PHOTOMETRIC AND SPECTRAL REDSHIFT DISTRIBUTIONS FROM CATALOGS, 
AND ON THE CMB INHOMOGENEITY IN PLANCK AND GRB\,021004 FIELD}
\label{sec:z_0_56_clustering:Sokolov_en}

\begin{figure*}
    \onelinecaptionstrue \captionstyle{normal} \setcaptionmargin{5mm}
    \includegraphics[width=1\columnwidth]{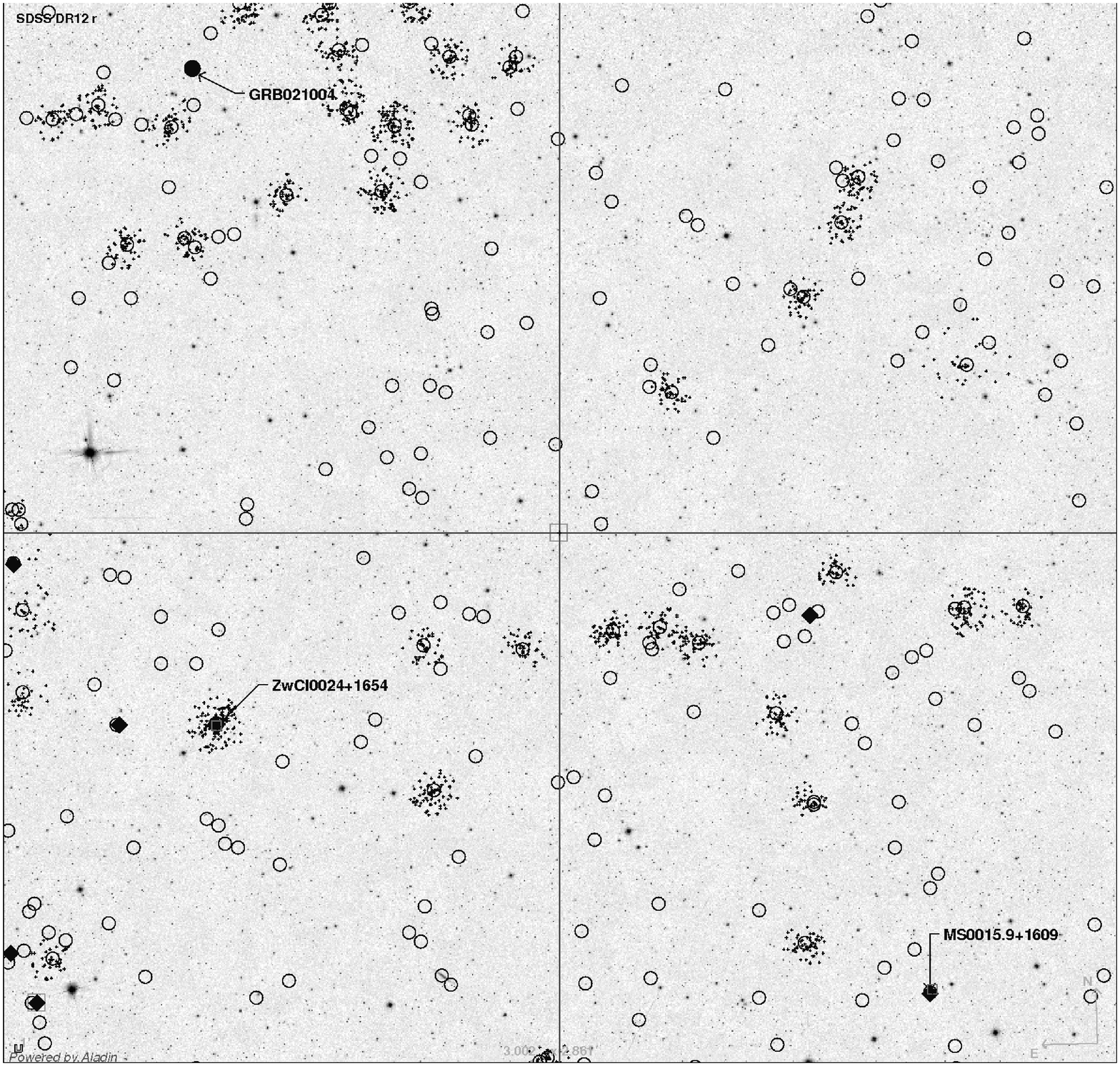}
     \caption{Area of $3\degr\times3\degr$ size encompassing the locations of
the gamma-ray burst GRB\,021004 (marked with filled circle) and two
known rich clusters---CL\,0024+1654 ($z=0.390$) and MS\,0015.9+1609
($z=0.541$). Other galaxy clusters from different catalogs (248 in
total) are marked. The empty  circles correspond to centers of galaxy
clusters  from the catalogs redMaPPer~\cite{22:Sokolov_en},
NOCS~\cite{28:Sokolov_en}, WHL~\cite{30:Sokolov_en,27:Sokolov_en} and 
the works by Saulder et al.~\cite{25:Sokolov_en}, Oguri~\cite{29:Sokolov_en} and
Tempel et al.~\cite{31:Sokolov_en}. The diamonds mark the rich
clusters from Abell and Zwicky catalogs~\cite{29:Sokolov_en}. Small
clumping crosses mark the galaxies---members of clusters from the
catalog redMaPPer~\cite{24:Sokolov_en,39:Sokolov_en}.}
        \label{3x3:Sokolov_en}
\end{figure*}

The peak near $z\approx0.56$ in distribution of redshifts for
galaxy clusters near the GRB\,021004 position was also tested
considering catalog data of both spectral and photometric
redshifts. We selected objects in $3\degr\times3\degr$ area
centered to the coordinates 
\mbox{${\rm RAJ}=00^{\rm h}22^{\rm m}44^{\rm s}$}, 
${\rm DecJ}=+17\degr40\arcmin58\arcsec$
(Fig.~\ref{3x3:Sokolov_en}). The total list of catalogs used for
this purpose (with their depths and the number of objects found)
is given below:

\begin{list}{$\bullet$}{
\setlength\leftmargin{5mm} \setlength\topsep{1mm}
\setlength\parsep{-0.5mm} \setlength\itemsep{2mm} }%
    \item {redMaPPer DR8 cluster catalog}~\cite{24:Sokolov_en,39:Sokolov_en}:
    the ca\-ta\-log depth is $m_i< 21\fm0$ for photometric redshifts, 
    44 clusters with $z_{\rm ph}$;
    \item {Group catalogues of the local Universe}~\cite{25:Sokolov_en}:
    the depth of the catalog  is $m_r< 17\fm77$ for photometric redshifts, 
    44 clusters with $z_{\rm sp}$;
    \item {Rich Clusters of Galaxies}~\cite{26:Sokolov_en}:
    the catalog depth is $m_B < 23^{\rm m}$ for spectral redshifts, 
    two clusters with $z_{\rm sp}$;
    \item {Newly rich galaxy clusters identified in
    \mbox{SDSS-DR12}}~\cite{27:Sokolov_en}:
    the depth of the catalog is $m_r < 21\fm5$ for photometric redshifts, 
    128 clusters with $z_{\rm sp}$ and $z_{\rm ph}$;
    \item {Northern Optical Cluster Survey} III~\cite{28:Sokolov_en}:
    the ca\-ta\-log depth is $21\fm5$, $21\fm0$ and $20\fm3$ in $g, r$ 
    and $i$  for photometric redshifts, 101 clusters with $z_{\rm ph}$;
    \item {Richness of galaxy clusters}~\cite{29:Sokolov_en}:
    the depth of the catalog is $m_i< 21\fm0$ for photometric redshifts,
74 clusters with $z_{\rm ph}$.
\end{list}

\begin{figure*}
    \onelinecaptionstrue \captionstyle{normal} \setcaptionmargin{5mm}
    \includegraphics[width=1\columnwidth]{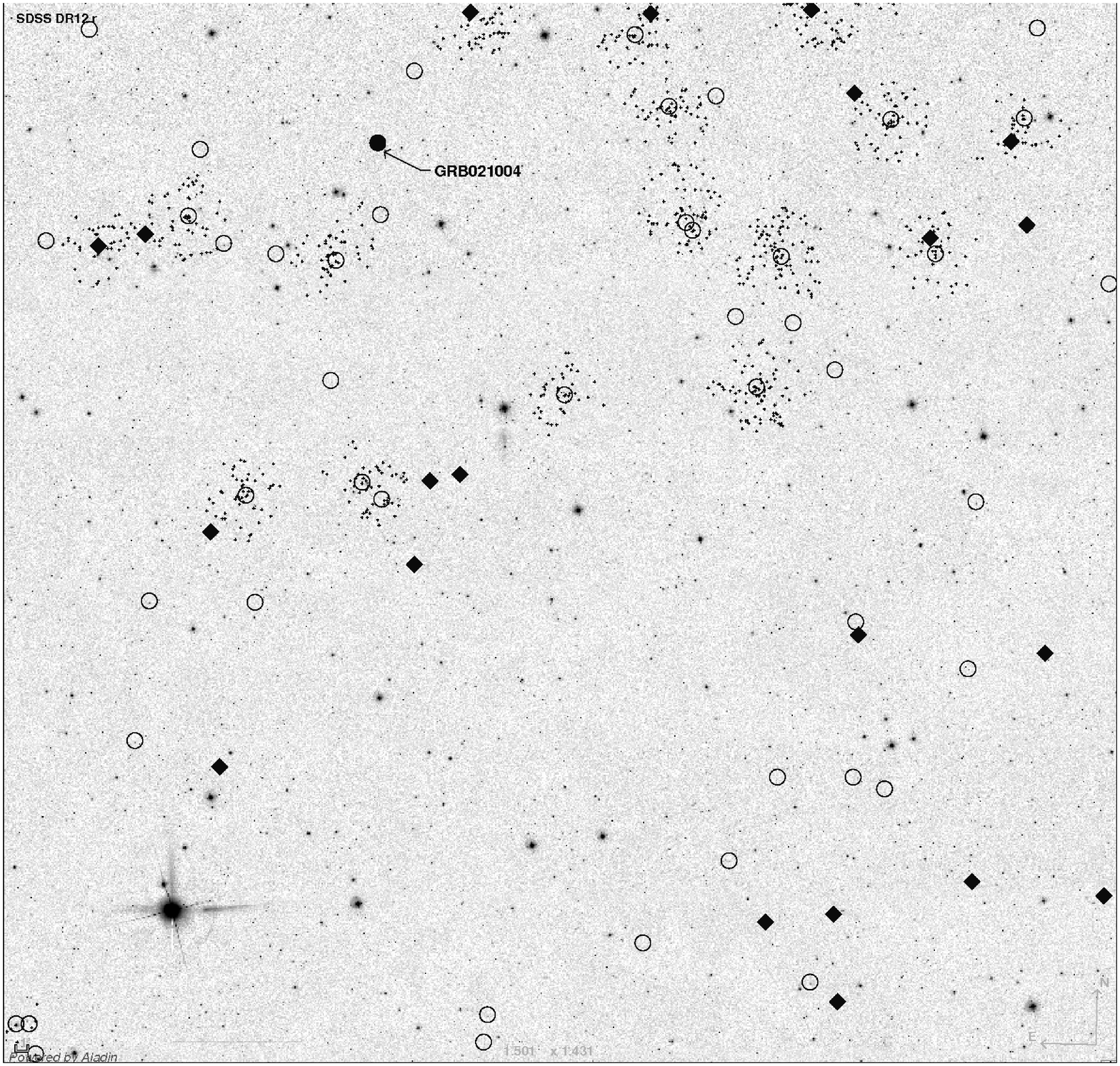}
\caption{The north-east part of the region shown in Fig.~\ref{3x3:Sokolov_en}
(see its upper left square). The area size is about $1\fdg5\times1\fdg5$. 
The notations are the same as in Fig.~\ref{3x3:Sokolov_en}.}
    \label{northeast:Sokolov_en}
\end{figure*}

The mosaic halftone image presented in Fig.~\ref{3x3:Sokolov_en}
is obtained thanks to the SDSS SAS\footnote{Science Archive
Server, \protect\url{https://dr12.sdss.org/mosaics}} from the
co-adding of 690 fields in the band $r$. The field contains two
X-ray neighboring galaxy clusters CL0024+1654 ($z=0.390$) and
MS\,0015.9+1609 ($z=0.541$) at angular distances of about
one-two degrees from GRB\,021004.
Figure~\ref{northeast:Sokolov_en} shows only the north-east part
(upper left square) of the whole region shown in
Fig.~\ref{3x3:Sokolov_en}, namely, the region of about
\mbox{$1\fdg5\times1\fdg5$} in size including the
GRB\,021004 location.
\begin{figure*}
    \onelinecaptionstrue \captionstyle{normal} \setcaptionmargin{5mm}
    \includegraphics[width=0.7\columnwidth]{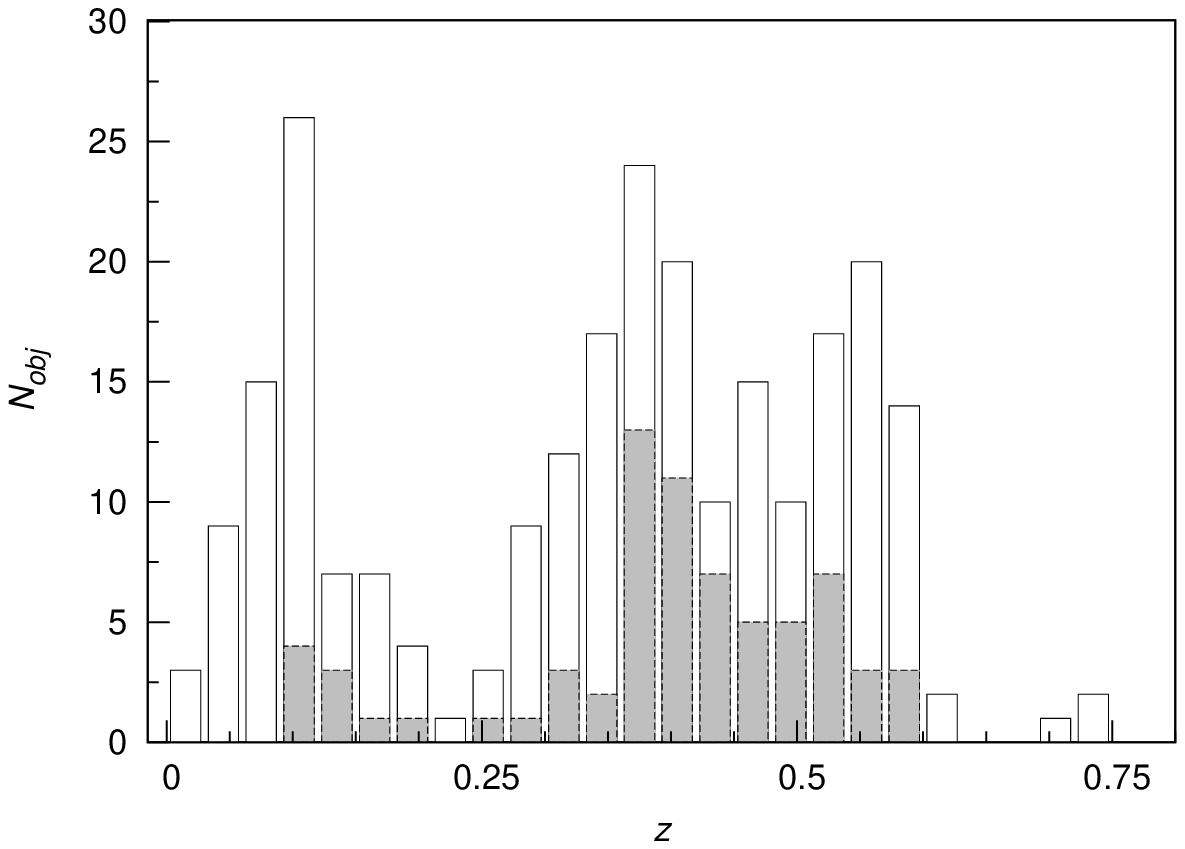}
\caption{The histograms of redshift distributions for clusters in the
region near GRB\,021004 location (Fig.~\ref{northeast:Sokolov_en}):
photometric and spectral redshifts, and  only spectral ones --light
and grey bins respectively.}
    \label{z-hist:Sokolov_en}
\end{figure*}

In Fig.~\ref{z-hist:Sokolov_en} the histograms of photometric
and spectral redshift distributions for galaxy clusters in this
$1\fdg5\times1\fdg5$ size region near GRB\,021004 location
(Fig.~\ref{northeast:Sokolov_en}) are shown. The redshift
distribution of clusters in this region concentrates towards at
least three peaks when using these new catalogs:
redMaPPer~\cite{24:Sokolov_en}, NOCS~\cite{28:Sokolov_en},
WHL~\cite{30:Sokolov_en,27:Sokolov_en}.

\begin{figure*}
    \setcaptionmargin{5mm}\onelinecaptionstrue
    \centerline{
      \vbox{
        \includegraphics[width=0.6\columnwidth]{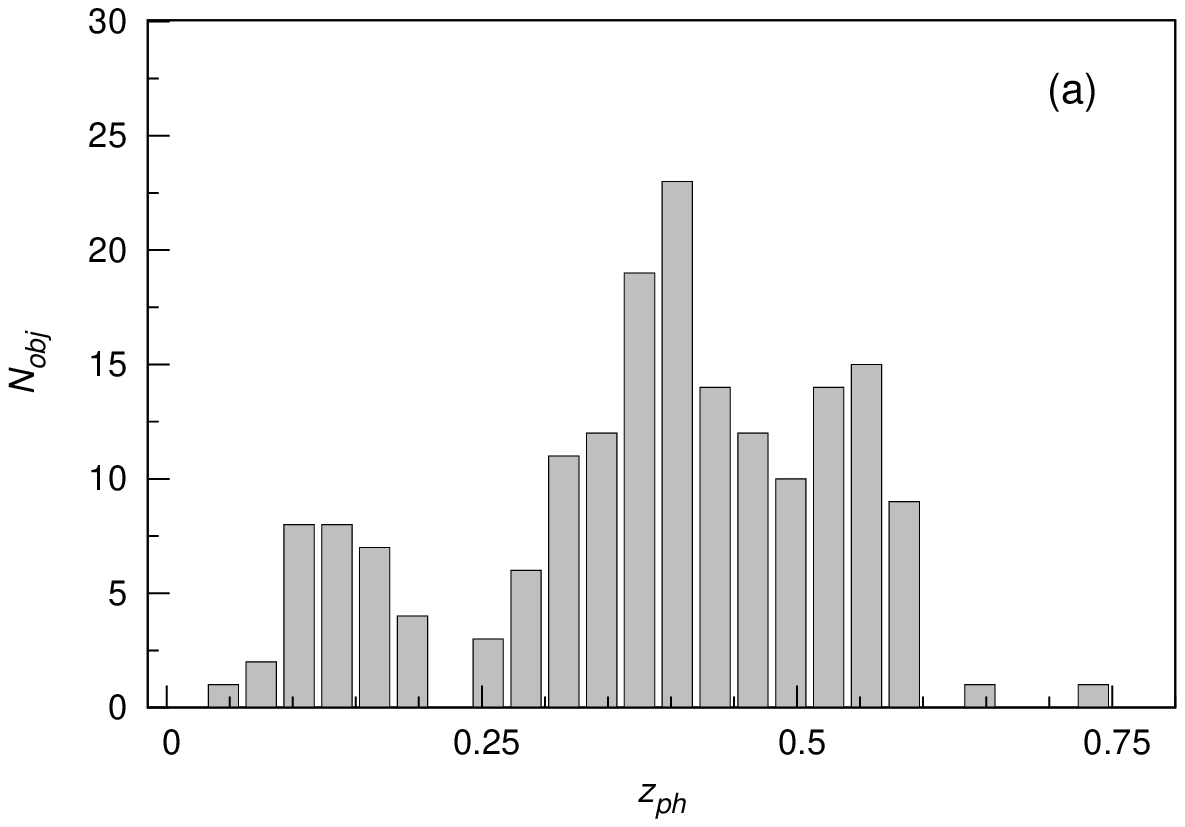}
        \includegraphics[width=0.6\columnwidth]{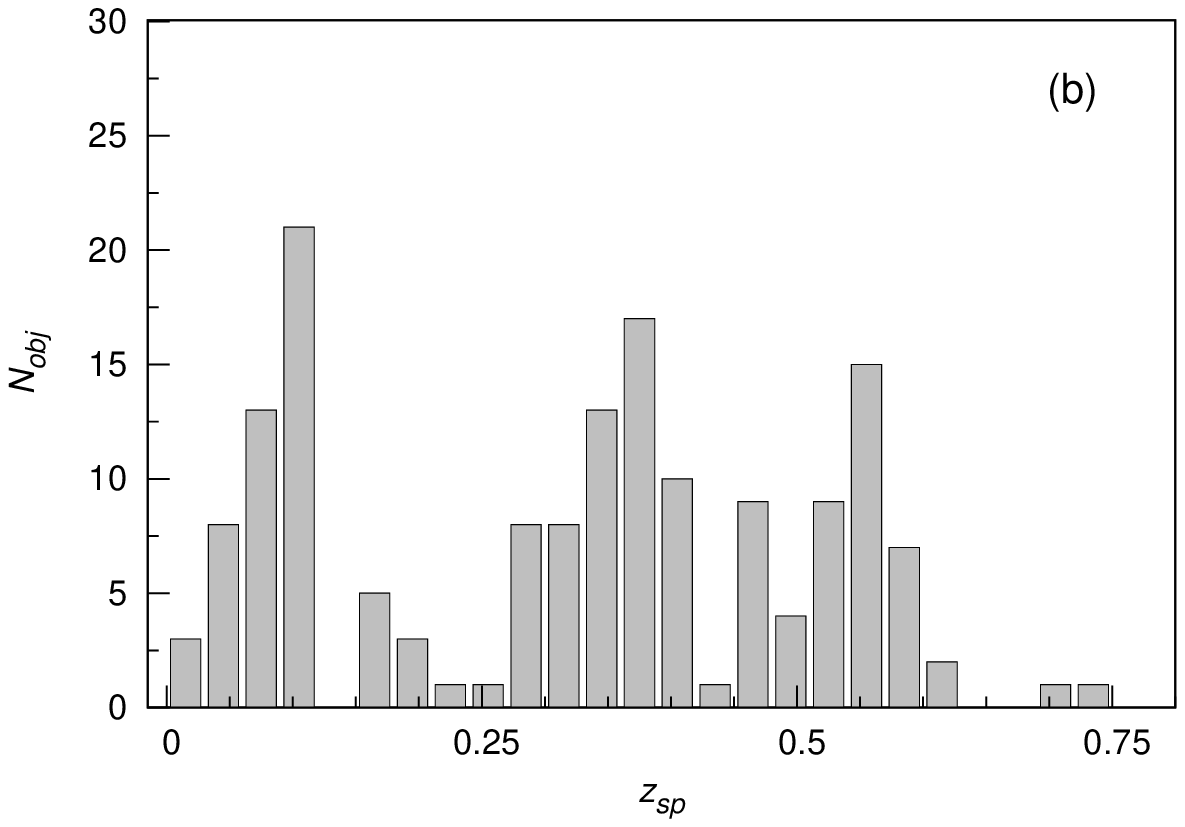}
        \includegraphics[width=0.6\columnwidth]{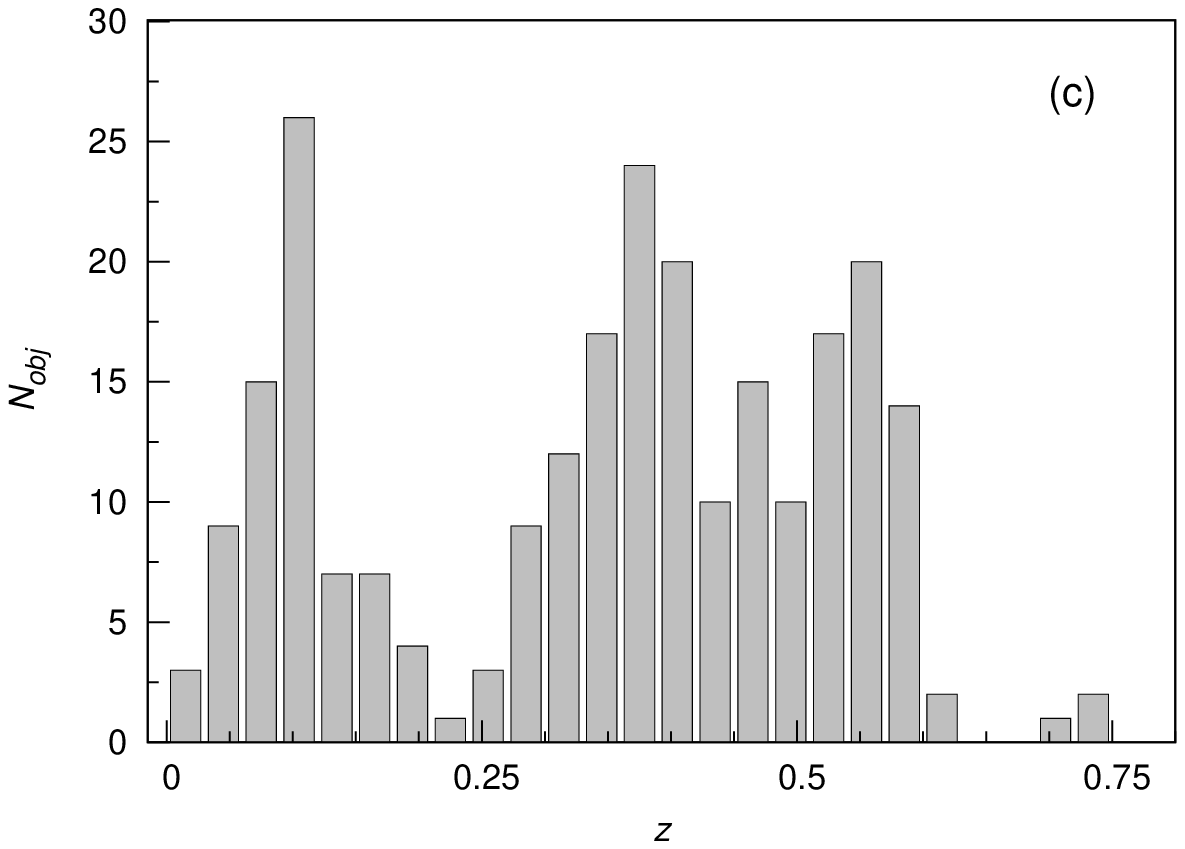}
}
}
\caption{Differential counts of redshifts for galaxy clusters from
six catalogs mentioned in Section 4 considering a region of
about $3\degr\times3\degr$ size centered at 
$RAJ=5\degr68$ ($00^{h}22^{m}43^{s}$), 
$DecJ=17\degr68$ ($17\degr40\arcmin48\arcsec$): 
(a)~photometric redshifts (180 clusters), (b)~spectral ones (160 clusters) and
(c)~spectral and photometric redshifts (248 clusters).}
    \label{z-histograms:Sokolov_en}
\end{figure*}

Figures~\ref{z-histograms:Sokolov_en}a--\ref{z-histograms:Sokolov_en}c
show the histograms of  differential counts of redshifts for
galaxy clusters found in the whole region of about 
$3\degr\times3\degr$ size with the center at 
${\rm RAJ}\!=\!5\degr68~(00^{\rm h}22^{\rm m}43^{s})$,
 \mbox {${\rm DecJ}\!=\!17\degr68$}~$(17\degr40\arcmin48\arcsec)$ 
(Fig.~\ref{3x3:Sokolov_en}). All six
catalogs (eight tables from Vizier database) were used---see the
beginning of the Section.

Thus the photometric redshifts estimated from the deep BTA $BVRI$
photometry turn out to be quite acceptable, as follows from
comparison with data of these catalogs---we see the same peak with 
$z\approx0.56$ in the redshift distribution for galaxy clusters near
the GRB\,021004 position from catalog data in the region with
approximately $1\fdg5\times1\fdg5$~size and even more---in the
area with size of about $3\degr\times3\degr$ at least.

{\it A practical remark.} A peculiarity detected when building
these distributions should be noted. If the catalog data for this
field (centered at the position $5\degr68;+17\degr68$, with the box
size of $3\degr\times3\degr$) are selected directly from the VizieR database,
then the query result actually somewhat exceeds the given size:
the distance between the outermost objects in the right ascension
is about \mbox{$11192\arcsec=3^\circ6\arcmin32\arcsec$}. Thus, in
fact, VizieR gives a field of $3\fdg1\times3\fdg0$ size, i.e.
more objects are found. This leads to the fact that histograms
analogous to those shown in Fig.~\ref{z-histograms:Sokolov_en}
and built from catalogs will slightly differ due to inclusion of
these objects, which should be taken into account with repeating
these $z$ distributions.

\begin{figure*}
    \onelinecaptionstrue \captionstyle{normal} \setcaptionmargin{5mm}
    \includegraphics[width=0.9\columnwidth]{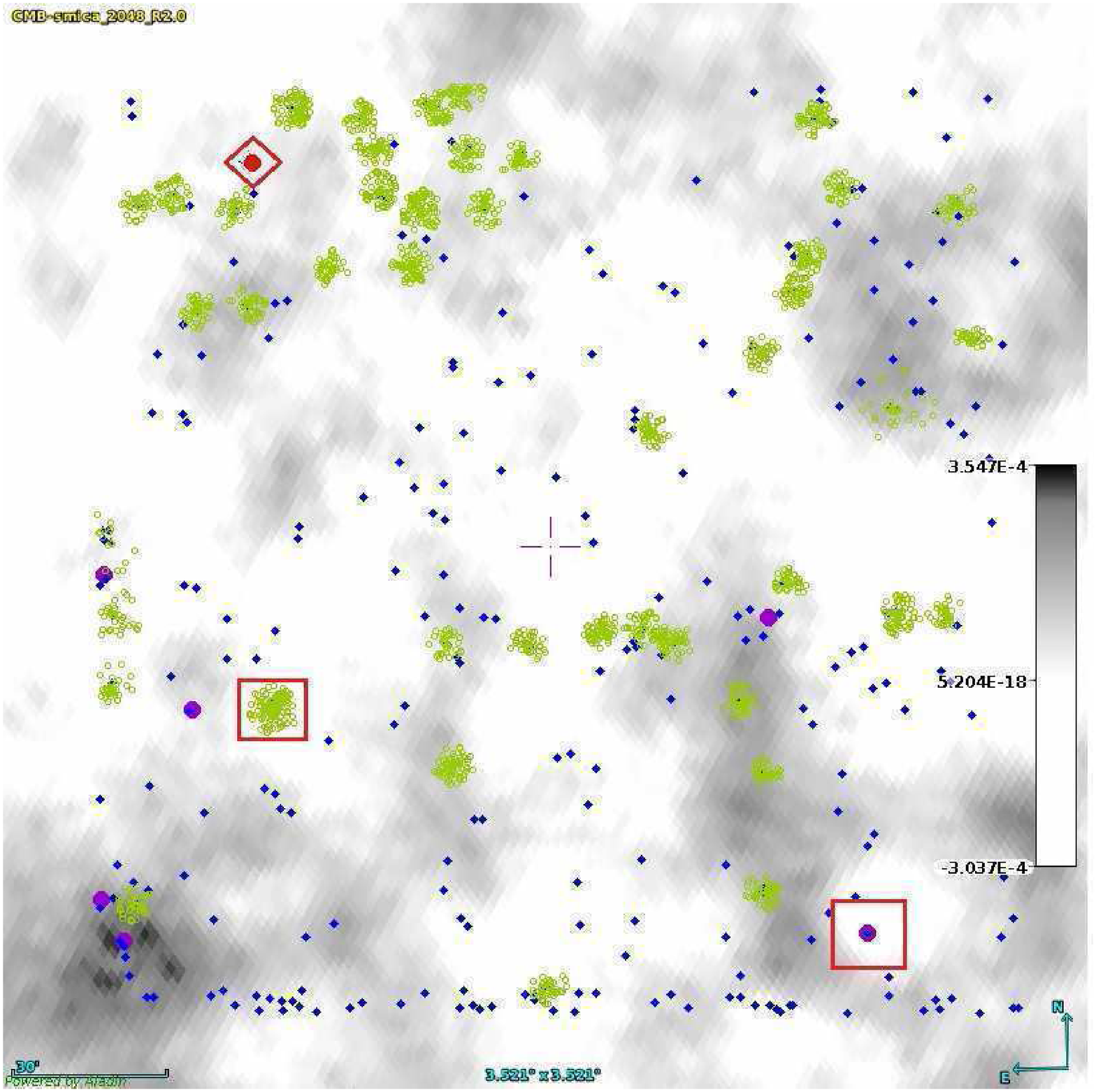}
    \caption{The same region as in Fig.~\ref{3x3:Sokolov_en}, but the size is
$3\fdg5\times3\fdg5$. The image is a CMB map (SMICA) from
archives of the {\it Planck} mission. Two known rich clusters
CL\,0024+1654 ($z=0.390$) and MS\,0015.9+1609 ($z=0.541$) and
other galaxy clusters from the catalogs are marked. The circle
(outlined by a diamond) marks the GRB\,021004 location. The full
circles show the location of Abell and Zwicky clusters, diamonds
show the location of clusters from the catalogs
redMaPPer~\cite{24:Sokolov_en}, NOCS~\cite{28:Sokolov_en},
WHL~\cite{30:Sokolov_en,27:Sokolov_en} and the works by
Saulder et al.~\cite{25:Sokolov_en},
Oguri~\cite{29:Sokolov_en} and Tempel
et~al.~\cite{31:Sokolov_en}.}
    \label{CMBmap:Sokolov_en}
\end{figure*}

In Fig.~\ref{CMBmap:Sokolov_en} the optical image of
Fig.~\ref{3x3:Sokolov_en} is overlaid on the {\it Planck} CMB (SMICA)
map\footnote{\url{http://pla.esac.esa.int/pla}}. The image
contains locations of GRB\,021004 and two galaxy clusters
CL\,0024+1654, and MS\,0015.9 with X-ray source in them.  These
two clusters are placed near white areas in
Fig.~\ref{3x3:Sokolov_en} corresponding to minima in the cosmic
microwave background (CMB) flux, what may be connected with the
Sunyaev--Zeldovich effect observed in such X-ray clusters.

Thus, using SDSS~DR12 data, we studied also a larger, of
approximately $3\degr\times3\degr$, region including these two
X-ray clusters. The depth of this survey is about $22^{\rm m}$,
and its catalog includes photometrical $z$ (their distribution is
shown in Fig.~\ref{z-histograms:Sokolov_en}a) and determinations
of spectral redshifts for sufficiently bright objects of the
survey (the distribution is presented in
Fig.~\ref{z-histograms:Sokolov_en}b). In the whole region (of
about $3\degr\times3\degr$), as well as in the region with
GRB\,021004 location ($\approx1\fdg5\times1\fdg5$ in
size) shown in Fig.~\ref{northeast:Sokolov_en}, the redshift
distribution of galaxy clusters also demonstrates an obvious peak
at $z \sim 0.56$ (see Figs.~\ref{z-hist:Sokolov_en}
and~\ref{z-histograms:Sokolov_en}a, \ref{z-histograms:Sokolov_en}b,
and~\ref{z-histograms:Sokolov_en}c) for photometrical and
spectral redshifts.

As it is seen in Figs.~\ref{northeast:Sokolov_en} and
\ref{z-histograms:Sokolov_en}, two X-ray clusters CL\,0024+1654
($z\!=\!0.390$) and MS\,0015.9+1609 ($z\!=\!0.541$) are located
very close in $z$ to two peaks at $z\!\sim\!0.4$ and
$z\!\sim\!0.56$ in the whole region of about
$3\degr\times3\degr$ size centered at \mbox {${\rm
RAJ}=5.68$}~$(00^{\rm h}22^{\rm m}43^{\rm s})$, ${\rm
DecJ}=17.68$~$(17\degr40\arcmin48\arcsec)$.

\section{SIZE ESTIMATION OF INHOMOGENEITY IN DISTRIBUTION 
OF GALAXY CLUSTERS WITH THE PEAK NEAR $z \approx 0.56$}
\label{sec:size_estimation:Sokolov_en}
\begin{figure}
    \onelinecaptionstrue \captionstyle{normal} \setcaptionmargin{5mm}
    \includegraphics[width=0.7\columnwidth]{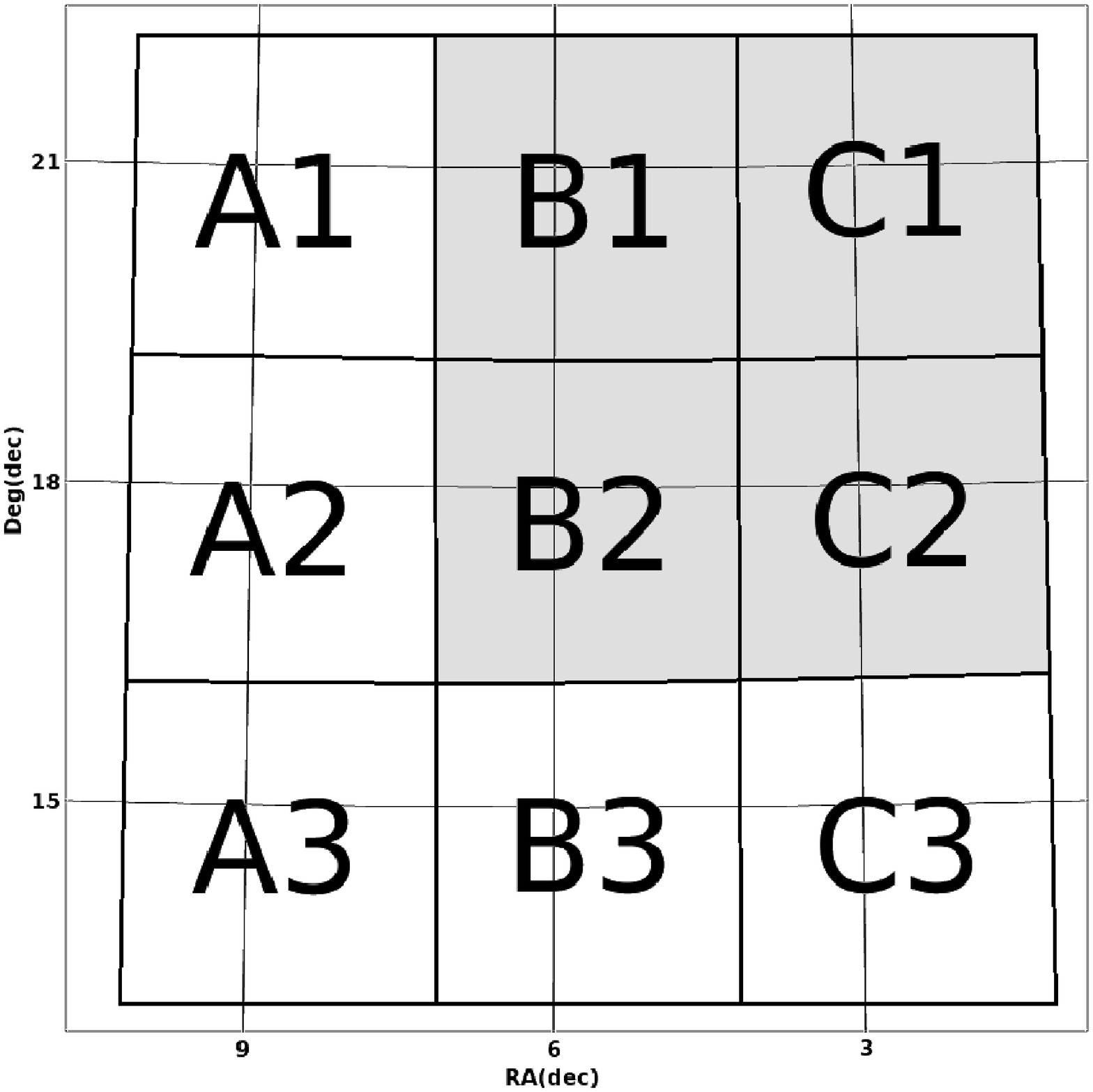}
    \caption{The layout of eight regions surrounding the central region B2.}
    \label{layout:Sokolov_en}
\end{figure}

To study the spatial behavior of the
overdensity excess of field galaxies near $z\sim0.56$ around the
GRB\,021004 position we considered also the redshift distributions in
adjacent areas. Relative position of the regions under consideration
is shown in the layout of Fig.~\ref{layout:Sokolov_en}. The central part B2 of
about $3\degr\times3\degr$ size practically coincides with the
region shown in Fig.~\ref{3x3:Sokolov_en}. It is surrounded by eight fields of
identical size.

\begin{figure*}
    \setcaptionmargin{5mm} \onelinecaptionstrue \captionstyle{normal}
\centerline{
\vbox{
\hbox{
\includegraphics[width=0.5\columnwidth]{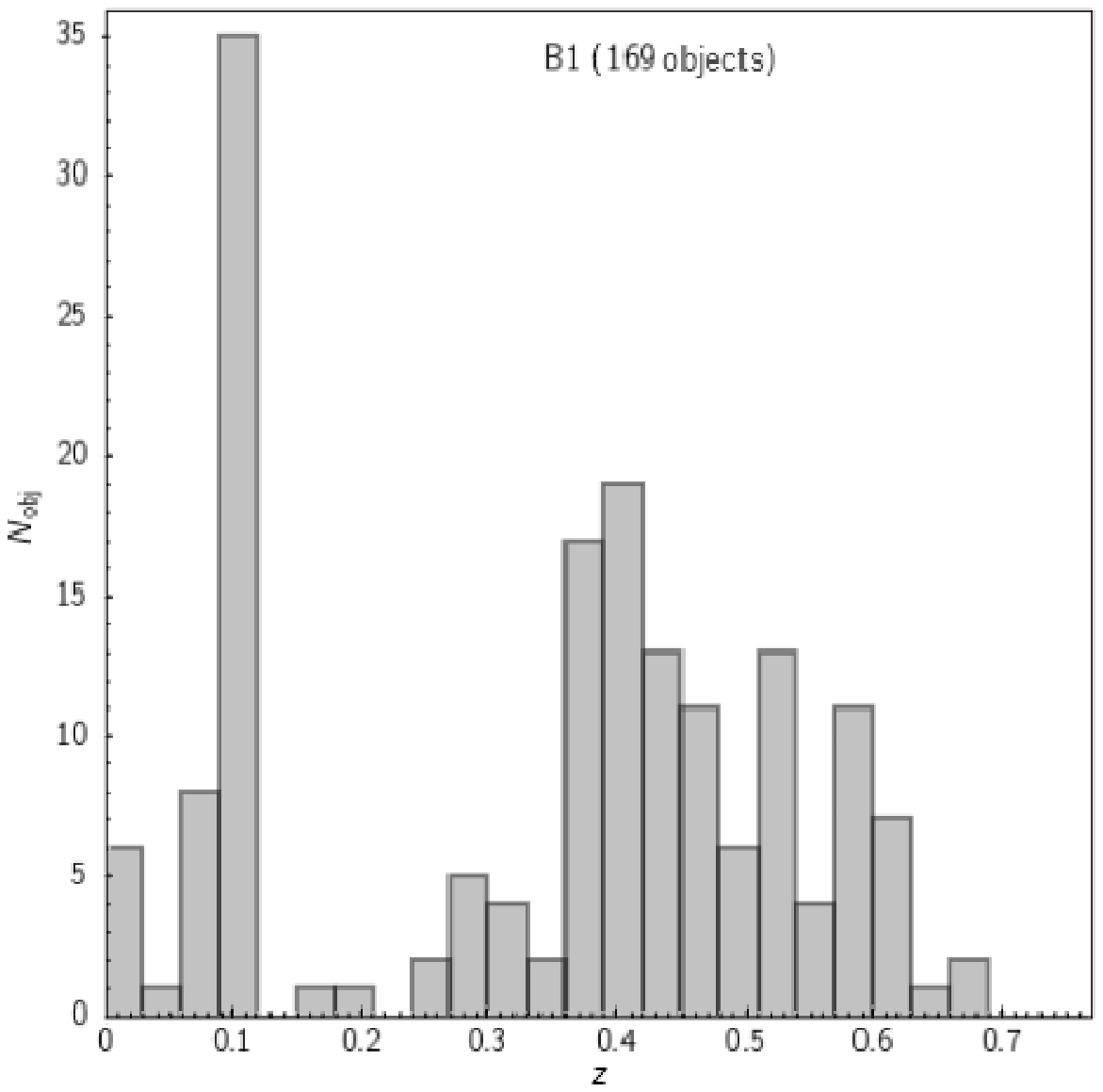}
\includegraphics[width=0.5\columnwidth]{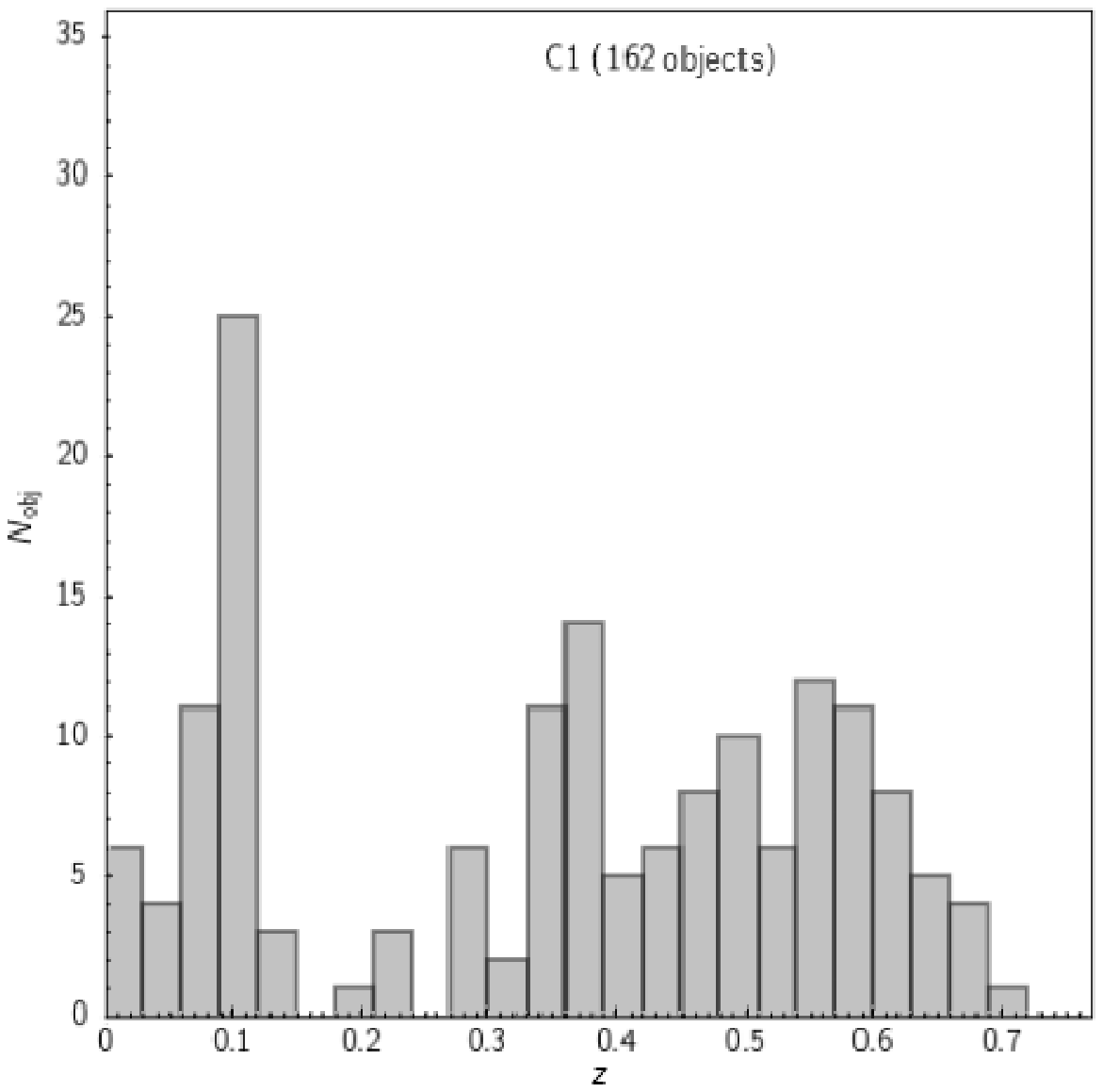}
}
\hbox{
\includegraphics[width=0.5\columnwidth]{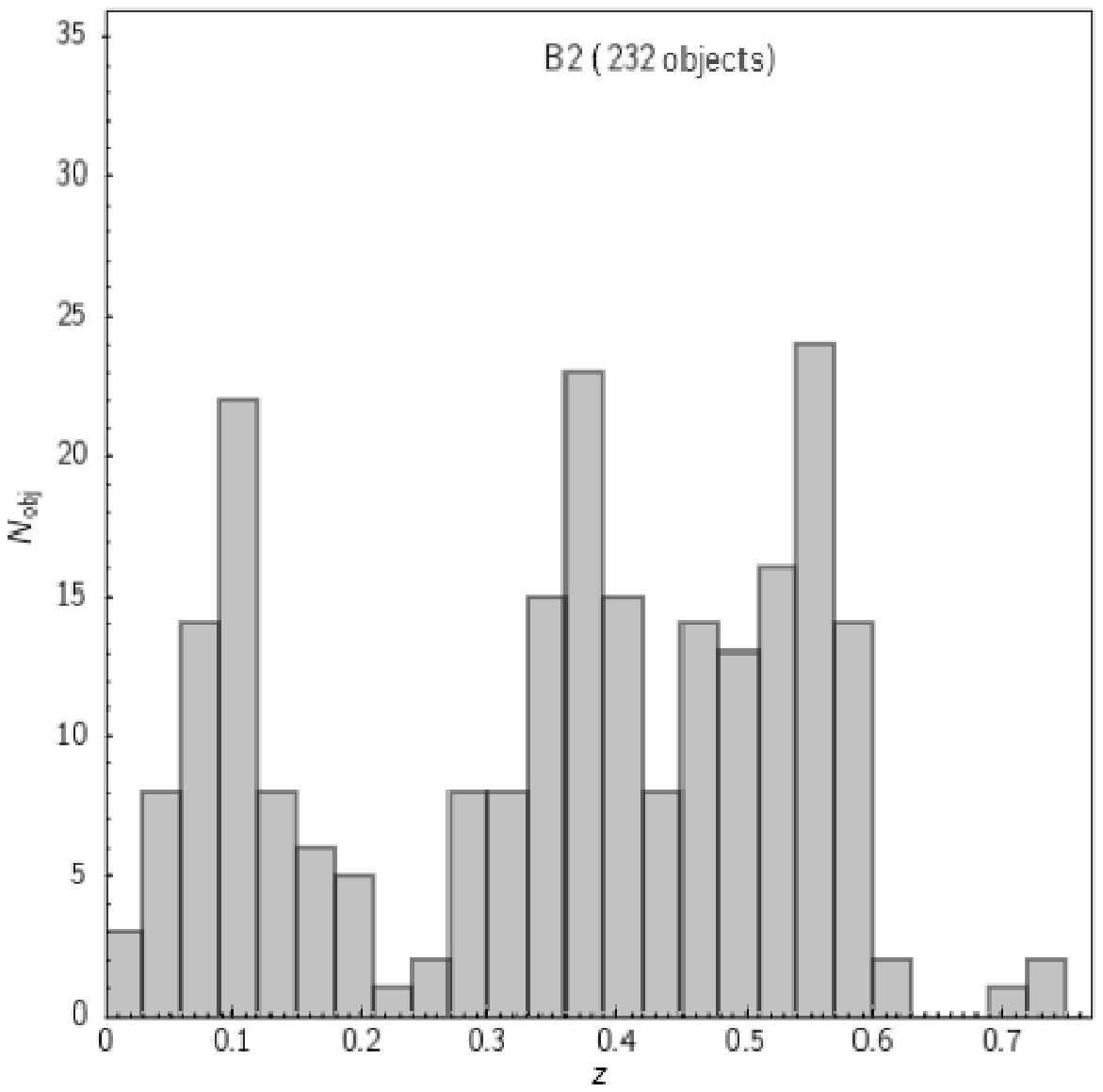}
\includegraphics[width=0.5\columnwidth]{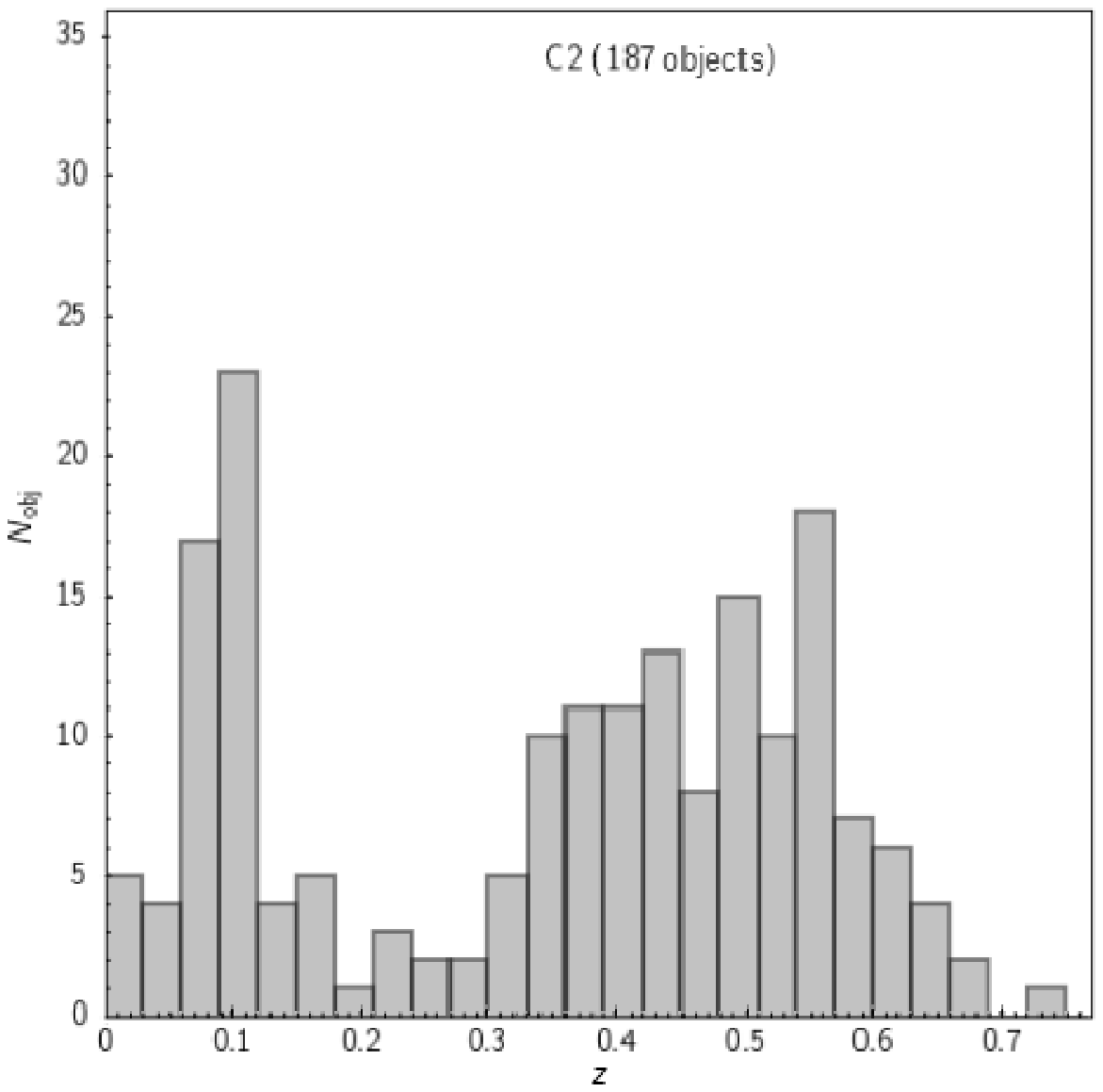}
}
}
}
\caption{Histograms of $z$ distributions for four fields
        (B1, B2, C1, C2) where the number of detected galaxy clusters
        near $z\sim0.56$ is the highest---the north-west square in
        the layout of Fig.~\ref{layout:Sokolov_en}. At the top of each
        histogram there is the name of the field for which it was built
        and the number of objects with known $z$ from catalogs in it.}
    \label{B1-C2-z-hist:Sokolov_en}
\end{figure*}
Using the same six above-mentioned catalogs
\mbox{\cite{24:Sokolov_en,39:Sokolov_en,25:Sokolov_en,26:Sokolov_en,27:Sokolov_en,28:Sokolov_en,29:Sokolov_en}}
we selected all galaxy clusters with known average (photometric
and spectral) redshifts in 8 regions surrounding the central
region B2. Applying the method as given in
Section~\ref{sec:z_0_56_clustering:Sokolov_en} for the central
region B2 (with the highest peak at about $z\approx0.56$), we
have built $z$ distributions of galaxy clusters analogous to those
shown in Fig.~\ref{z-histograms:Sokolov_en} for each of these
eight regions.

Figure~\ref{B1-C2-z-hist:Sokolov_en} presents histograms of
$z$-distributions only for four of nine fields (B1, B2, C1, C2),
where the number of detected galaxy clusters near $z\sim0.56$ is
the largest (the north-west  square in
Fig.~\ref{layout:Sokolov_en}). As is seen, the peak near $z\sim0.56$ 
in these distributions is only seen in the B2, C1 and C2 regions.
In the region B1 the number of galaxy clusters with this $z$ is less.

\begin{figure*}
    \onelinecaptionstrue \captionstyle{normal} \setcaptionmargin{5mm}
    \includegraphics[width=0.9\columnwidth]{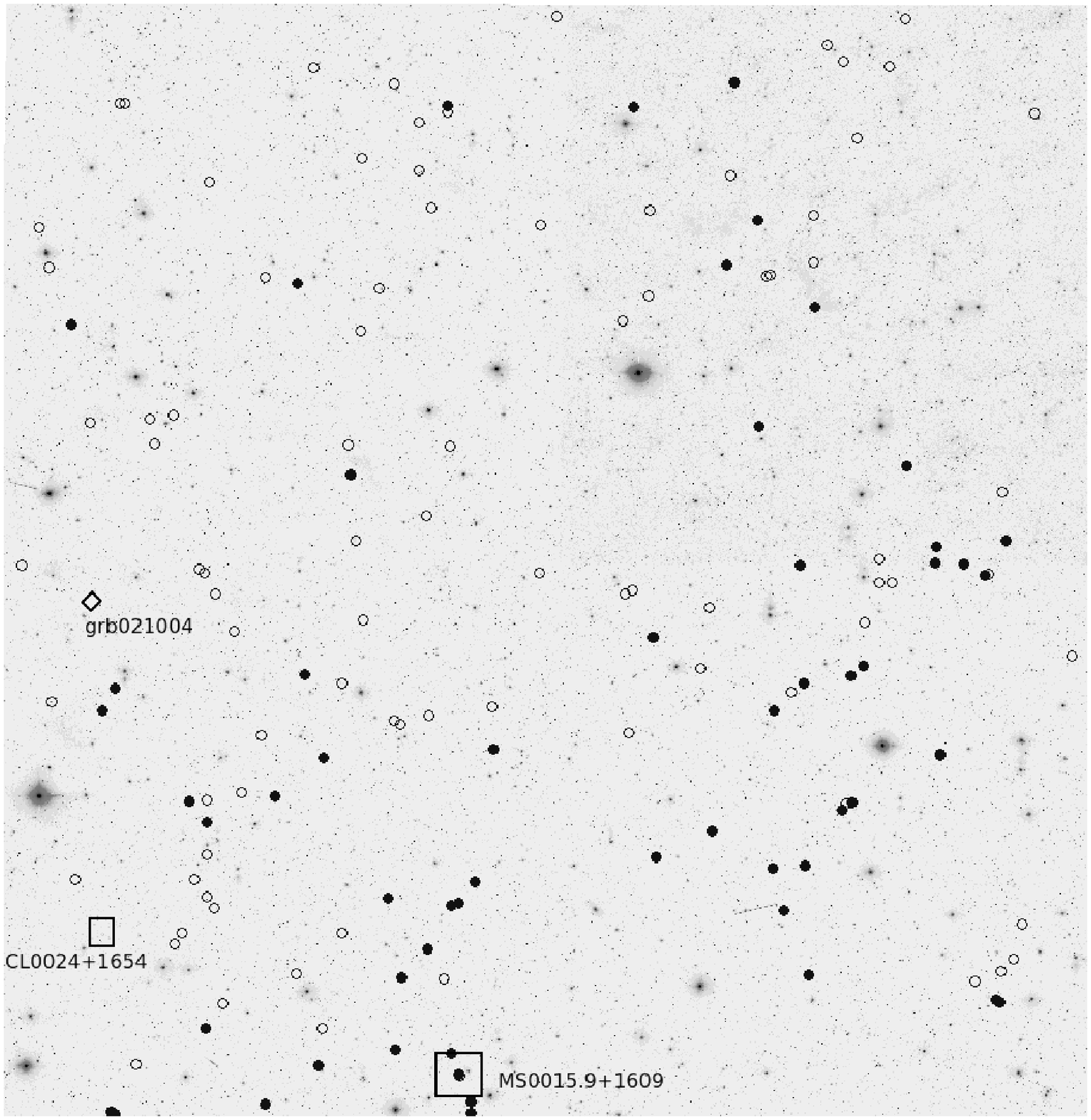}
    \caption{Direct images of four fields B1, B2, C1, C2 corresponding
    to the layout in Fig.~\ref{CMBmap:Sokolov_en} and histograms
    in Fig.~\ref{B1-C2-z-hist:Sokolov_en}---a square of size
    $6\degr\times6\degr$. The catalog objects---the centers of
    galaxy clusters---with redshifts near the peak at $z\approx0.56$
    are marked. The filled circles denote the objects with redshifts in the
    range $0.54<z<0.57$ corresponding to the peak bin. The unfilled
    circles are objects from the adjacent bins $0.51<z<0.54$ and $0.57<z<0.6$.}
    \label{B1-C2-mosaic:Sokolov_en}
\end{figure*}

Figure~\ref{B1-C2-mosaic:Sokolov_en} shows the direct images of four fields B1, B2, C1, C2 
corresponding to the layout of Figure~\ref{layout:Sokolov_en} and
histograms in Fig.~\ref{B1-C2-z-hist:Sokolov_en} (a square of size
\mbox{$6\degr\times6\degr$}) where the number of detected galaxy
clusters near \mbox{$z\sim0.56$} is the highest. In
Fig.~\ref{B1-C2-mosaic:Sokolov_en} the catalog objects---the
centers of galaxy clusters---with redshifts near the peak at $z\approx0.56$ are marked.

Thus, from catalog data (certainly, with allowance made for their
depths) one can estimate the size of the overdensity excess as
about $6\degr$--$8\degr$ in distribution of galaxy clusters with
the peak near \mbox{$z\approx0.56$}.

\section{DISCUSSION AND CONCLUSIONS} % (fold)
\label{sec:discussion_and_conclusions:Sokolov_en}

The main idea of this work is to reveal the signatures of field
galaxies clustering in the GRB\,021004 line of sight. And we
tests for reliability any available
signatures of the clustering in the direction and near
the GRB\,position.
\begin{list}{}{
\setlength\leftmargin{5mm} \setlength\topsep{1mm}
\setlength\parsep{-0.5mm} \setlength\itemsep{2mm} }
    \item[$\bullet$] The first signature is the BTA GRB\,021004 deep
    field (with a size of about $4\arcmin \times 4\arcmin$)
    photometric redshift distribution with a peak near
    to $z\sim0.56$ estimated from BTA multicolor  $BVRI$
    photometry in the GRB\,direction.
    \item[$\bullet$] The second signature is the Mg\,II
    $\lambda\lambda2796,2803$\r{A}\r{A} absorption doublet at $z\approx0.56$
    in the GRB\,021004 afterglow VLT/UVES spectra obtained
    for the afterglow and GRB host galaxy.
    \item[$\bullet$] The third signature is the galaxy clustering
    in a larger ($\approx3\degr\times3\degr$) area around
    GRB\,021004 with an effective peak near $z\sim0.56$ for spectral
    and photometric redshift distributions from
    the SDSS and BOSS as a part of SDSS\mbox{-}III.
    \item[$\bullet$] And the fourth signature may be some inhomogeneity
    in the {\it Planck} field, what may be due to the
    Sunyaev--Zeldovich effect for galaxy clusters near the GRB\,021004 position.
    \item[$\bullet$] Also, from data of six catalogs the size of the
    whole inhomogeneity in distribution of galaxy clusters with the
    peak near $z\approx0.56$ was estimated to be about $6\degr$--$8\degr$.
\end{list}
In all redshift distributions presented (Figs.~\ref{z-hist:Sokolov_en},
\ref{z-histograms:Sokolov_en} and
\ref{B1-C2-z-hist:Sokolov_en}) the peak at $z\approx0.1$ is
obviously distinguished, which corresponds to the well known Sloan
Great Wall as a complex of superclusters with a huge size of about
$7^{\rm h}$~\cite{1:Sokolov_en,2:Sokolov_en}. 
In the same figures, a peak with $z\approx0.4$ is also seen, which can
indicate an inhomogeneity in distribution of galaxy clusters near
the direction to GRB\,021004 with $z\sim0.4$. The same
supercluster can also include the X-ray cluster CL\,0024+1654 with
$z=0.390$ (see Figs.~\ref{3x3:Sokolov_en}
and~\ref{northeast:Sokolov_en}). The X-ray cluster
MS\,0015.9+1609 with $z=0.541$ in
Fig.~\ref{B1-C2-mosaic:Sokolov_en}, is in the range $0.54<z<0.57$, 
corresponding to the peak
bin in Fig.~\ref{z-histograms:Sokolov_en} with $z\sim0.56$.

The GRB\,021004 host galaxy with $z_{\rm ph}=2.225$ is also 
in Fig.~\ref{z-distribution:Sokolov_en}
in distribution of photometric redshifts. This $z_{\rm ph}$ was
measured from BTA $BVRI$ photometry for the GRB\,host galaxy with
$B=24\fm434\pm0\fm132$, $V=24\fm006\pm0\fm099$, $R=24\fm174\pm0\fm154$,
$I=23\fm437\pm0\fm170$. Correspondingly, the spectroscopic
measurements of redshift for GRB\,021004 are $z_{\rm sp}=2.3295$~\cite{19:Sokolov_en} and
$2.3304$~\cite{20:Sokolov_en}. 
Thus, the photometric z ph of this galaxy measured by us with BTA
corresponds to the spectral z with the identical error of about 10%.

Here it should be emphasized that the obtaining of data for
estimation of photometric redshift can be done at once for a large
sample of galaxies and down to limit depths (see
Fig.~\ref{mag-err:Sokolov_en}) in the whole GRB field when
studying properties of the same host
galaxies~\cite{14:Sokolov_en}. At present, the photometric
redshifts are already widely used in all modern investigations of
galaxy clusters at different $z$, which we also use here (see
Sections~\ref{sec:z_0_56_clustering:Sokolov_en} and
\ref{sec:size_estimation:Sokolov_en}). The new
paper~\cite{45:Sokolov_en} deals with application of the
photometric redshift method for 1227 galaxies in the Hubble Ultra
Deep field down to 30th magnitude (down to \mbox{$F775W=30$}) in
the $0.4<z<1.5$ range. Here we would like also to draw
attention to another recent work~\cite{47:Sokolov_en}, in which
the matter is on photometry in five bands, and the photometric
redshift estimations use the galaxy morphology information. This
paper shows that by adding galaxy morphological parameters to full
$u, g, r, i, z$-photometry, only mild improvements are obtained,
while the gains are substantial in cases where fewer passbands are
available. For instance, the combination of $g,r,z$-photometry and
morphological parameters almost fully recovers the metrics of
five-band photometric redshifts.

When analyzing our BTA $BVRI$ data we have considered a redshift range up to
$z=4$, though the number of objects with $z>1$, which are
bright enough for such analysis, becomes less and less, see
Fig.~\ref{z-distribution:Sokolov_en}. Correspondingly, the
uncertainty in estimation of photometric redshifts of such (more
and more distant) objects becomes larger according
to~\cite{16:Sokolov_en} where the $z$ estimation errors were
studied specially. The limits of catalogs and BTA $BVRI$ data are
also mentioned at the end of
Section~\ref{sec:redshift_distribution:Sokolov_en}.

Thus, the results strongly depends on the photometry precision, see
Fig.~\ref{mag-err:Sokolov_en}. Here we have chosen approximately
10\% for the objects under investigation. Correspondingly, for
objects at redshifts more than~1 the needed exposure times are
larger than those indicated in
Table~\ref{Tab:fields:Sokolov_en}. Besides, here the infrared
filters can be already needed also. But since here we are
interested in the range of intermediate redshifts with $z < 0.7$,
for which there are already catalog data on photo-z (see at the
end of Section~\ref{sec:redshift_distribution:Sokolov_en} +
references therein and the beginning of
Section~\ref{sec:z_0_56_clustering:Sokolov_en}), we use these
photo-$z$ (+~spectro-$z$) from catalogs to test our calculations.

Here it should be said also about spectroscopy in the GRB afterglow
direction, about our results and (in this connection) about new
observational tasks.

A comprehensive study based on a spectroscopic data of 73~GRB
afterglows discussing the GRB/quasar overdensity excess of field
galaxies around sight-lines was reported~\cite{18:Sokolov_en},
included the above-mentioned Mg~II $\lambda\lambda2796,2803$\r{A}\r{A}
doublet.

If this excess of intervening systems is real, it should be
possible to find an excess of GRB field galaxies around GRB
positions, although a preliminary study has revealed no anomalous
clustering of galaxies (in comparison with distribution of quasar)
at the estimated median redshift of about 0.3 around GRB line of
sights~\cite{18:Sokolov_en}. Furthermore, it has been proposed
that the majority of short-duration GRBs in early-type galaxies
will occur in clusters and three such relationships have been
already found~\cite{33:Sokolov_en}.

It is also well proven that long-duration GRBs are associated with
the core collapse of very massive
stars~\cite{34:Sokolov_en,35:Sokolov_en,36:Sokolov_en}. 
Similarly to core collapse supernovae, the collapse of massive stellar
iron cores results in the formation of a compact object (collapsar),
accompanied by the high-velocity ejection of a large fraction of
a progenitor star mass at relativistic speed producing a series of
internal shocks giving rise to the GRB itself~\cite{37:Sokolov_en}.

It is known~\cite{40:Sokolov_en} that measurements of the X-ray
surface brightness of galaxy clusters (including such as
CL\,0024+1654 with $z=0.390$ and MS\,0015.9+1609 with $z=0.541$) 
can be used to estimate the angular diameter and distance
to these structures. Namely, the determination of distance to
supernovae and gamma-ray bursts resulting from collapse of compact
objects of stellar mass becomes the main observational task in
determining a basic parameter: the total energy release related to
such events. The collapse of the massive stellar cores maybe
connected with the quark phase transition in the compact objects,
which leads to neutrino, gravitational and photon signals from the
core collapse supernovae (like SN\,1987A) and GRBs. It is also
obvious that for low and intermediate redshifts, the sky
distribution of electromagnetic and neutrino signals associated
with the core collapses can be
non-isotropic~\cite{38:Sokolov_en}, showing the clustering of
galaxies in which the formation of compact objects occurs due to
evolution of massive stars.

In conclusion we would like also to emphasize here that since GRBs
are detected at more and more distant cosmological distances with
redshifts up to 9.2~\cite{10:Sokolov_en}, this poses additional new
questions which are of outmost importance for observational
cosmology.  What are the redshifts at which the sky distribution of
GRBs becomes homogeneous?  And what are the redshifts where such
bursts (which are related now with collapse of compact objects of
stellar mass) are unobservable already?

\begin{acknowledgments}
The authors  are grateful Yu.~Baryshev for fruitful discussion and
valuable comments.
\end{acknowledgments}

%\bibliographystyle{AstroBull}
%\bibliography{Sokolov}

\begin{thebibliography}{99}
%%
% by VNK
%%
%\newcommand{\enquote}[1]{``#1''}
%\providecommand{\selectlanguage}[1]{\relax}

\bibitem{1:Sokolov_en}
R.~K.~{Sheth} and A.~{Diaferio}, \mnras\ \textbf{417}, 2938
(2011).

\bibitem{2:Sokolov_en}
M.~{Einasto}, H.~{Lietzen}, M.~{Gramann}, et~al., \aap\
\textbf{595}, A70  (2016).

\bibitem{3:Sokolov_en}
H.~{Lietzen}, E.~{Tempel}, L.~J.~{Liivam{\"a}gi}, et~al., \aap\
\textbf{588}, L4 (2016).

\bibitem{4:Sokolov_en}
R.~G.~{Clowes}, K.~A.~{Harris}, S.~{Raghunathan}, et~al., \mnras\
\textbf{429},  2910 (2013).

\bibitem{37:Sokolov_en}
T.~{Piran}, Phys. Reports \textbf{314}, 575 (1999).

\bibitem{6:Sokolov_en}
L.~G.~{Bal{\'a}zs}, Z.~{Bagoly}, J.~E.~{Hakkila}, et~al., \mnras\
\textbf{452},  2236 (2015).

\bibitem{41:Sokolov_en}
I.~{Horv{\'a}th}, J.~{Hakkila}, and Z.~{Bagoly}, \aap\ \textbf{561},
L12 (2014).

\bibitem{7:Sokolov_en}
M.~L.~{Khabibullina}, O.~V.~{Verkhodanov}, and V.~V.~{Sokolov},
\ab\ \textbf{69}, 472 (2014).

\bibitem{8:Sokolov_en}
C.~A.~{Meegan}, G.~J.~{Fishman}, R.~B. {Wilson}, et~al., \nat\
\textbf{355}, 143  (1992).

\bibitem{9:Sokolov_en}
D.~{Fargion}, arXiv:1408.0227 (2014).

\bibitem{11:Sokolov_en}
S.~{Raghunathan}, R.~G.~{Clowes}, L.~E.~{Campusano}, et~al.,
\mnras\  \textbf{463}, 2640 (2016).

\bibitem{12:Sokolov_en}
S.-F.~S.~{Chen}, R.~A.~{Simcoe}, P.~{Torrey}, et~al., \apj\
\textbf{850}, 188  (2017).

\bibitem{10:Sokolov_en}
V.~{Sudilovsky}, J.~{Greiner}, A.~{Rau}, et~al., \aap\
\textbf{552}, A143  (2013).

\bibitem{42:Sokolov_en}
M.~{Arabsalmani}, P.~{M{\o}ller}, D.~A.~{Perley}, et~al., \mnras\
\textbf{473},  3312 (2018).

\bibitem{13:Sokolov_en}
T.~A.~{Fatkhullin}, A.~A.~{Vasil'ev}, and V.~P.~{Reshetnikov},
\alet \textbf{30}, 283 (2004).

\bibitem{14:Sokolov_en}
V.~V.~{Sokolov}, T.~A.~{Fatkhullin}, A.~J.~{Castro-Tirado},
et~al., \aap\  \textbf{372}, 438 (2001).

\bibitem{18:Sokolov_en}
G.~E.~{Prochter}, J.~X.~{Prochaska}, and S.~M.~{Burles}, \apj\
\textbf{639}, 766  (2006).

\bibitem{19:Sokolov_en}
S.~D.~{Vergani}, P.~{Petitjean}, C.~{Ledoux}, et~al., \aap\
\textbf{503}, 771  (2009).

\bibitem{20:Sokolov_en}
A.~J.~{Castro-Tirado}, P.~{M{\o}ller}, G.~{Garc{\'{\i}}a-Segura},
et~al., \aap\  \textbf{517}, A61 (2010).

\bibitem{44:Sokolov_en}
V.~L.~{Afanasiev} and A.~V.~{Moiseev}, \alet \textbf{31}, 194
  (2005).

\bibitem{43:Sokolov_en}
Y.~N.~{Parijskij}, O.~P.~{Zhelenkova}, P.~{Thomasson}, et~al., %in \emph{EAS
%   Publications Series}, Edited by A.~J. {Castro-Tirado}, J.~{Gorosabel}, and
%   I.~H. {Park} (2013), \emph{EAS Publications Series}, vol.~61, pp. 439--442.
EAS Publ. Ser. {\bf 61}, 439 (2013).

\bibitem{15:Sokolov_en}
E.~{Bertin} and S.~{Arnouts}, \aaps\ \textbf{117}, 393 (1996).

\bibitem{21:Sokolov_en}
W.~A. {Baum}, %in \emph{Problems of Extra-Galactic Research}, %Edited by G.~C.
%   {McVittie} (1962), \emph{IAU Symposium}, vol.~15, p. 390.
IAU Symp. {\bf 15}, 390 (1962).

\bibitem{22:Sokolov_en}
M.~{Bolzonella}, J.-M.~{Miralles}, and R.~{Pell{\'o}}, \aap\
\textbf{363}, 476  (2000).

\bibitem{23:Sokolov_en}
D.~{Calzetti}, L.~{Armus}, R.~C.~{Bohlin}, et~al., \apj\
\textbf{533}, 682  (2000).

\bibitem{16:Sokolov_en}
Y.~V.~{Baryshev}, I.~V.~{Sokolov}, A.~S.~{Moskvitin}, et~al., \ab\
\textbf{65}, 311 (2010).

\bibitem{24:Sokolov_en}
E.~S.~{Rykoff}, E.~{Rozo}, D.~{Hollowood}, et~al., \apjs\
\textbf{224}, 1  (2016).

\bibitem{39:Sokolov_en}
E.~S.~{Rykoff}, E.~{Rozo}, M.~T.~{Busha}, et~al., \apj\
\textbf{785}, 104  (2014).

\bibitem{25:Sokolov_en}
C.~{Saulder}, E.~{van Kampen}, I.~V.~{Chilingarian}, et~al., \aap\
\textbf{596},  A14 (2016).

\bibitem{26:Sokolov_en}
G.~O.~{Abell}, H.~G.~{Corwin},~Jr., and R.~P.~{Olowin}, \apjs\
\textbf{70}, 1  (1989).

\bibitem{27:Sokolov_en}
Z.~L.~{Wen} and J.~L.~{Han}, \apj\ \textbf{807}, 178 (2015).

\bibitem{28:Sokolov_en}
R.~R.~{Gal}, P.~A.~A.~{Lopes}, R.~R.~{de~Carvalho}, et~al., \aj\
\textbf{137},  2981 (2009).

\bibitem{29:Sokolov_en}
M.~{Oguri}, \mnras\ \textbf{444}, 147 (2014).

\bibitem{30:Sokolov_en}
Z.~L.~{Wen}, J.~L.~{Han}, and F.~S.~{Liu}, \apjs\ \textbf{199}, 34
(2012).

\bibitem{31:Sokolov_en}
E.~{Tempel}, A.~{Tamm}, M.~{Gramann}, et~al., \aap\ \textbf{566}, A1
(2014).

\bibitem{45:Sokolov_en}
J.~{Brinchmann}, H.~{Inami}, R.~{Bacon}, et~al., \aap\ \textbf{608},
A3 (2017).

\bibitem{47:Sokolov_en}
J.~Y.~H. {Soo}, B.~{Moraes}, B.~{Joachimi}, et~al., \mnras\
\textbf{475}, 3613  (2018).

\bibitem{33:Sokolov_en}
M.-S. {Shin} and E.~{Berger}, \apj\ \textbf{660}, 1146 (2007).

\bibitem{34:Sokolov_en}
H.~{Y{\"u}ksel} and M.~D.~{Kistler}, Phys. Let. B \textbf{751}, 413  (2015).

\bibitem{35:Sokolov_en}
M.-H.~{Li} and H.-N.~{Lin}, \aap\ \textbf{582}, A111 (2015).

\bibitem{36:Sokolov_en}
T.~N.~{Ukwatta} and P.~R.~{Wo{\'z}niak}, \mnras\ \textbf{455}, 703 (2016).

\bibitem{40:Sokolov_en}
R.~F.~L.~{Holanda}, V.~C.~{Busti}, L.~R.~{Cola{\c c}o}, et~al., J.
Cosmology Astroparticle Phys.  \textbf{8}, 055 (2016).

\bibitem{38:Sokolov_en}
A.~{Gomboc}, Contemporary Physics \textbf{53}, 339 (2012).

\end{thebibliography}

\end{document}